\documentclass[preprint]{aastex62}

\shorttitle{Spectral Synthesis Analysis of SDSS J0850$+$4451}
\shortauthors{Leighly et al.}

\begin{document}


\title{The z=0.54 LoBAL Quasar SDSS J085053.12+445122.5: I. Spectral
  Synthesis Analysis Reveals a Massive Outflow
    \footnote{Based on observations made with the NASA/ESA Hubble
      Space Telescope, obtained from the Data Archive at the Space
      Telescope Science Institute, which is operated by the
      Association of Universities for Research in Astronomy, Inc.,
      under NASA contract NAS 5-26555. These observations are
      associated with program \#13016.}}


\author{Karen M.\ Leighly}
\affiliation{Homer L.\ Dodge Department of Physics and Astronomy, The
  University of Oklahoma, 440 W.\ Brooks St., Norman, OK 73019}

\author{Donald M.\ Terndrup}
\affiliation{Homer L.\ Dodge Department of Physics and Astronomy, The
  University of Oklahoma, 440 W.\ Brooks St., Norman, OK 73019}
\affiliation{Department of Astronomy, The Ohio State University, 140
  W.\ 18th Ave., Columbus, OH 43210} 

\author{Sarah C.\ Gallagher}
\affiliation{The Centre for Planetary and Space Exploration, The
  University of Western Ontario}

\affiliation{The Rotman Institute of Philosophy, The University of
  Western Ontario}
\affiliation{Department of Physics \& Astronomy, The University of Western
  Ontario, London, ON, N6A 3K7, Canada}

\author{Gordon T.\ Richards}
\affiliation{Department of Physics, Drexel University, 3141 Chestnut Street,
  Philadelphia, PA 19104}

\author{Matthias Dietrich}
\altaffiliation{Deceased 19 July 2018}
\affiliation{Earth, Environment, and Physics, Worcester State University,
  Ghosh Science and Technology Center, Worcester, MA 01602}



%%%%%%%%%%%%%%%%%%%%%%%%%%%%%%%%%%%%%%%%%%%%%%%%

\begin{abstract}
We introduce {\it SimBAL}, a novel spectral-synthesis 
procedure that uses large grids of ionic column densities
generated by the photoionization code {\it Cloudy} and a Bayesian
model calibration to forward-model broad absorption line quasar
spectra.  We used {\it SimBAL} to analyze the {\it HST} COS 
spectrum of the low-redshift BALQ SDSS~J085053.12+445122.5.   {\it
  SimBAL} analysis yielded velocity-resolved information about the
physical conditions of the absorbing gas.  We found that the
ionization parameter and column density increase, and the covering
fraction decreases as a function of velocity. The total log column
density is 22.9 (22.4) [$\rm cm^{-2}$] for solar ($Z=3Z_\odot$)
metallicity.   The outflow lies 1--3 parsecs from the central engine,
consistent with the estimated location of the torus.  The mass outflow 
rate is 17--$28\rm \, M_\odot yr^{-1}$, the momentum flux is consistent
with $L_{Bol}/c$, and the ratio of the kinematic to bolometric
luminosity is  0.8--0.9\%.   The outflow velocity is similar to the
escape velocity at the absorber's location, and force multiplier
analysis indicates that part of the outflow could originate in
resonance-line driving.  The location near the torus suggests that 
dust scattering may play a role in the acceleration, although the lack 
of reddening in this UV-selected object indictes a relatively
dust-free line of sight. The low accretion rate ($0.06 L_{\rm Edd}$)
and compact outflow suggests that SDSS~J0850$+$4451 might be 
a quasar past its era of feedback, although since its mass outflow is
about 8 times the accretion rate, the wind is likely integral to the
accretion physics of the central engine.
\end{abstract}

\keywords{quasars: absorption lines --- quasars: individual (SDSS J085053.12+445122.5)}

\section{Introduction}\label{intro}

The optical and UV spectra of active galactic nuclei (AGN) and quasars
offer powerful diagnostics of the physical conditions of gas in the
vicinity of their central engines.  Powered by
photoionization, the broad  emission lines trace the kinematics of the
broad line region, likely dominated by Keplerian motions, while the
broad absorption lines trace the outflow.  A range of ionization
states are seen from a number of different ground- and excited-state
transitions. The broad emission lines are significantly
Doppler-broadened by the motion of the gas in the 
vicinity of the black hole, with characteristic velocity widths of 1000s of
$\rm km\, s^{-1}$,  making line blending a considerable
impediment to quantitative analysis.  Emission-line studies are further
compromised by the potential contributions to a single line from gas
with a wide range of illumination patterns (e.g., the illuminated side
of a cloud will emit differently than the back side of the cloud).
Absorption line studies are more straightforward because
only line-of-sight gas is important. Diagnostic power is
lost when lines are saturated and complicated by the fact that the
gas is known to partially cover the accretion disk, the source of the
continuum emission. While we know that the gas must be clumpy in order
to be dense enough to produce the emission and absorption that we see,
there is no clear understanding of the characteristic scale and
distribution of these  clumps, or of how they are formed and
maintained within the dynamic environment of the central engine. 

Nevertheless, emission and absorption lines in quasars provide insight
into fundamentally important phenomena.
The central engine of quasars is not only a bright beacon in the
Universe, exhibiting a rich phenomenology, but it also may be
contributing to regulating the rate of star formation in the host
galaxy \citep[e.g.,][]{kp15}, thus contributing to the observed tight
correlation between black hole mass and the mass of the bulge
\citep[e.g.,][]{fm00,kh13}. Blue-shifted absorption lines, in particular,
provide strong evidence for powerful outflows, and may prove to be key
tracers of how accretion power may couple to the host galaxy's 
interstellar medium.

Early quantitative analysis of broad absorption line quasar spectra
focused on estimating the physical conditions of the gas, including
ionization parameter, column density, and metallicity as a 
way of understanding the acceleration mechanism and potential for
chemical enrichment of the intergalactic medium.  Initially, these
investigations proceeded with little concern for the width of the
absorption lines. \citet{arav01b} reported the analysis of the
low-redshift quasar PG~0946$+$301, which has a FWHM of the main
\ion{C}{4} component of about $8000\rm \, km\, s^{-1}$.  Once partial
covering was discovered to be important, investigators started working
on determining the covering fraction and nature of partial covering
\citep[e.g.,][]{hamann01,dekool02c}.  Around the same time, scientists
began to appreciate the diagnostic power of absorption lines with
easily-populated excited states for determining the density of the
outflows \citep[e.g.,][]{dekool01,dekool02a,dekool02b}.  An issue is
that many of the diagnostic pairs of lines (e.g.,
\ion{C}{2} at 1334.0 and 1335.7 \AA\/, \ion{S}{4} at 1062.7 and
1073.0 \AA\/) lie quite close together in wavelength
\citep[e.g.,][Fig.\ 15]{lucy14}, making blending a problem for lines
that are broad.      

The focus on using excited states to determine the outflow density,
and the accompanying problems with blending, means that much of the recent
work to determine the physical conditions of the outflowing gas using
spectroscopic  diagnostics has been done on objects with relatively narrow
lines. For example, HE~0238$-$1904 has several components with
velocity widths of $500\rm\, km\, s^{-1}$ \citep{arav13}.
FBQS~J0209$-$0438 shows an absorption system with overall width of
$600\rm\, km\, s^{-1}$ \citep{finn14}.  QSO~2359$-$1241 shows
\ion{Fe}{2} absorption from various velocity components ranging in
width from $<50\rm \,km\, s^{-1}$ to $\sim 100 \rm \, km\, s^{-1}$
\citep{bautista10}.  SDSS~J1106$+$1939 shows 
\ion{S}{4} absorption with width of $\sim 2900\rm \, km\, s^{-1}$,
while SDSS~J1512$+$1119 shows \ion{S}{4} absorption with width of
$\sim 250\rm \, km\, s^{-1}$ \citep{borguet13}.  Yet the population of
BAL quasars shows an enormous range of velocity widths. 
\citet{baskin15} report analysis of the \ion{C}{4} line from 1596 BAL
quasars taken from the SDSS DR7 quasar catalog \citep{shen11}.  The
distribution of \ion{C}{4} width peaks at around $2000\rm \, km\,
s^{-1}$ with a long tail to larger velocities.  A cumulative 
distribution shows that 25\% have velocity widths larger than $5200
\rm \, km\, s^{-1}$, while 10\% have velocity widths larger than
$7500\rm\, km\, s^{-1}$.  Limiting analysis to objects with narrow 
lines, or selecting out exactly those quasars with the strongest winds
that are most likely to be important for feedback, may limit our
understanding of quasar outflows. 

Another issue was revealed by \citet{lucy14}.  In that paper, we
analyzed the iron low-ionization broad absorption line quasar
(FeLoBAL) FBQS~J1151$+$3822.  We modeled the lines using a normalized
absorption-line template developed  from the \ion{He}{1}* absorption
lines \citep{leighly11}.  Following the procedure in the literature
that has been used by many authors \citep[e.g.,][]{moe09,   dunn10,
  borguet12,arav13, chamberlain15}, we fit the template to the spectrum in order to
estimate the apparent column densities of line complexes.  We compared
the measured column densities with column densities predicted by the
photoionization code {\it Cloudy} \citep{ferland13}, and used a figure
of merit to determine the best-fitting values of ionization parameter
$\log U$, density, and column density (parameterized as $\log N_H -
\log U$).  Our next step was novel.  To check our best-fit result, we
created a synthetic spectrum using the best-fitting parameters and
overlaid it on the observed spectrum \citep[Fig.\ 3b,][]{lucy14}. The
resulting fit was a very poor match to the observed spectrum,
potentially implying that the physical parameters derived using this
type of analysis  may be wrong.   

Clearly a new approach is necessary, and we can take inspiration from
work with other energetic systems.  Supernovae are another type of
astronomical object with broad absorption lines.  In these objects,
the lines can be so broad that identification of a feature can be
difficult.  Spectral synthesis codes have proven invaluable for both
line identification \citep[SYNOW,][]{branch05} and 
for analysis of the physical conditions in the outflow
\citep[PHOENIX,][]{hb99}. It stands to reason that a similar approach
may be useful for broad absorption line quasars.  

To that end, we introduce {\it SimBAL}, a spectral-synthesis
forward-modeling method for analyzing BAL quasar spectra.  
{\it SimBAL}, in essence, inverts the conventional method for
analyzing absorption lines.  Instead of fitting individual lines and
then comparing those measurements with {\it Cloudy} models, synthetic
spectra 
are constructed from {\it Cloudy} models and then compared with the
observed spectrum.  The spectral synthesis approach has several
advantages.  First, because we do not  need to identify individual
absorption lines, blending is not an issue, so the width of the line
no longer is an impediment to selection of quasars for analysis.
Second, the conventional analysis method outlined above focuses on the
lines that are observed, neglecting the important information provided
by lines that are {\it not} detected.  Since the spectral synthesis
approach models the whole spectrum, the information provided by absent
lines is used 
to constrain the solution.  Finally, we use a Markov Chain Monte Carlo
method in physical parameter space to compare the synthetic spectrum
with the observed spectrum.  This method allows us to harvest uncertainties
on the physical parameters from the posterior probability
distributions.  Along the way, we have also discovered that we can
map the physical parameters of the outflow (e.g., ionization 
parameter, column density, and covering fraction) as a function of
velocity, properties that may be important for constraining
acceleration models for the outflows.  With the physical properties of
the outflow in hand and a few assumptions, we can estimate mass
outflow rates, key for constraining the kinetic energy available for
quasar feedback on the host galaxy.

In this paper, we use {\it SimBAL} to analyze {\it HST} COS spectrum 
of the low redshift ($z=0.5422$) LoBAL quasar
SDSS~J085053.12$+$445122.5, hereafter referred to as SDSS~J0850+4451.
The observation and continuum model are described in
\S\ref{observations}.  A brief description of {\it SimBAL} is given in
\S\ref{simbal}. The absorption modeling and extraction of the physical
parameters of the outflow is described in
\S\ref{absorption_modeling}. The implications of our analysis are
discussed in \S\ref{discussion}, the summary of our principal results and
future development of {\it SimBAL} are discussed in
\S\ref{conclusions}, { and several potential systematic effects are
  discussed in an Appendix.}
Vacuum wavelengths are used throughout. Cosmological parameters used
depend on the context (e.g., when comparing with results from an older
paper), and are reported in the text.  

\section{Observations and Data Reduction}\label{observations}

\subsection{HST COS Observations}\label{cos}

SDSS~J0850$+$4451 was observed with {\it HST}
COS \citep{osterman11} using the G230L grating on 2013 May 12. The
goal of the observation was to obtain high signal-to-noise spectra
covering the major broad absorption lines including
\ion{P}{5}$\lambda\lambda 1118, 1128$ on the short wavelength end and
\ion{C}{4} on the long wavelength end.   Two
central wavelengths were used.  The 33,971-second exposure using the 
2950\AA\/ setting provided rest-frame coverage from 1080 to 1380\AA\/.
The 8,848-second exposure using the 3360\AA\/ setting provided
rest-frame coverage between 1370 and 1640\AA\/.  The ``x1dsum''
spectra were extracted from the pipeline fits files.  

The spectrum was corrected for Milky Way reddening using
$E(B-V)=0.024$ \citep{sf11}.  The spectrum was shifted to the rest
frame using a  cosmological redshift of $0.5422$.  This value was
estimated from the narrow [\ion{O}{3}] line in the SDSS spectrum, and
is between 0.5423,  the value determined by \citet{hewett10}, and
0.5414, the NED\footnote{https://ned.ipac.caltech.edu/} preferred
redshift. 

\subsection{Continuum Model}\label{cont_model}

\begin{figure*}[!t]
\epsscale{1.0}
\begin{center}
\includegraphics[width=4.5truein]{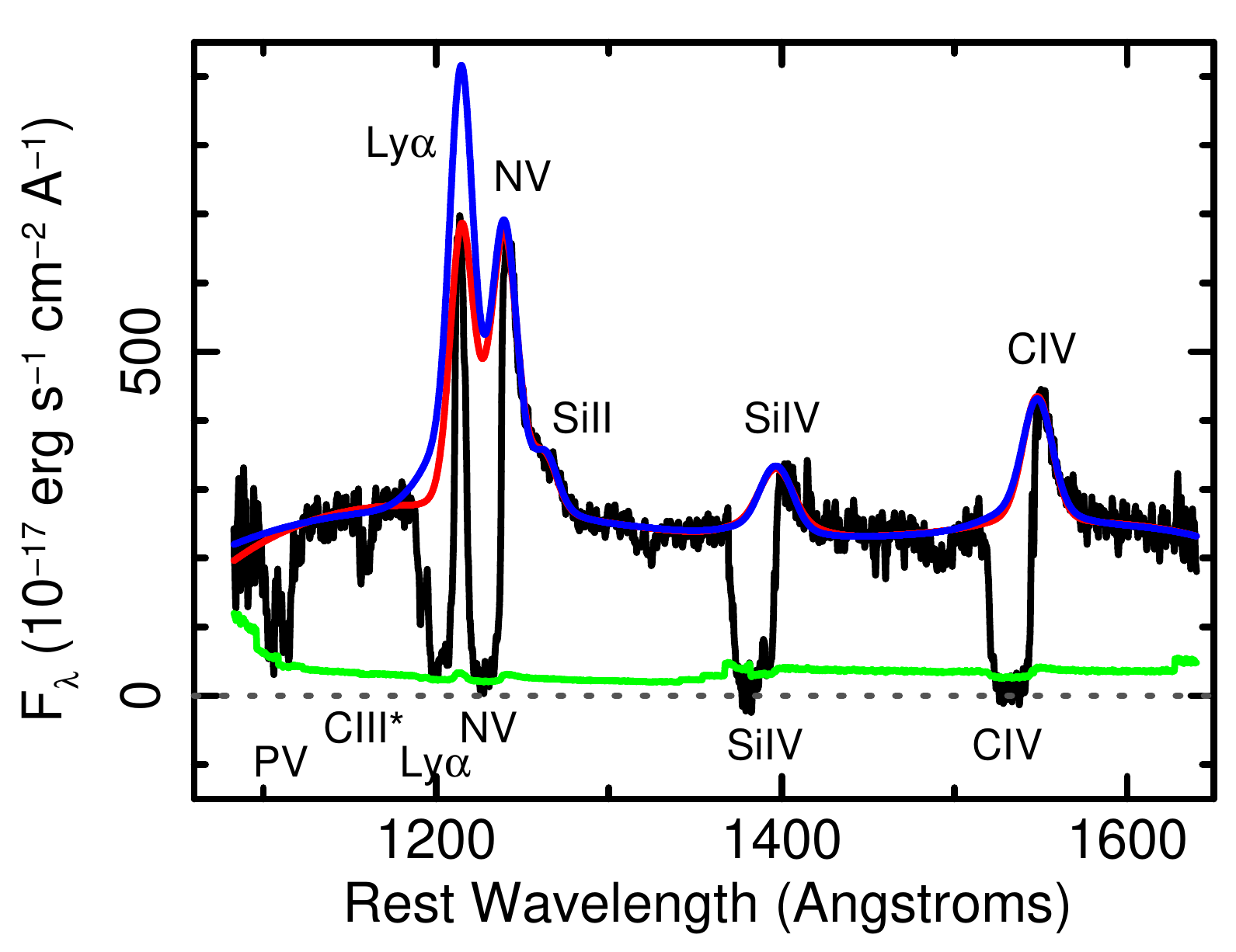}
\caption{The {\it HST} far-UV spectrum (black) and continuum models.
  The red line shows the model referred to as the ``first continuum''
  in the text, obtained assuming that the feature near 1212\AA\/ is 
  unabsorbed   continuum plus line emission.  The blue line   shows
  the model referred to as the ``second continuum'', obtained
  assuming that the Ly$\alpha$ emission line is partially 
  absorbed.   Principal emission lines are labeled above the spectrum,
  and principal absorption lines are labeled below.  See
  text for details.  The  statistical uncertainty in 
  the spectrum  is shown in green.  The mean signal to noise ratio in
  the spectrum  was 7.2, dropping to 2.2 shortward of $\sim 1100$
  \AA\/.   \label{fig3}}   
\end{center}
\end{figure*}

The {\it HST} spectrum shows rather prominent UV emission lines
typical of a broad-line AGN.  The spectrum appears to be similar to
the LBQS quasar composite spectrum \citep{francis91}.
\citet{leighly04} showed that a range of \ion{C}{4} profiles can be
modeled using an intermediate-width component at approximately the
rest wavelength and a broad and blueshifted component, with variable
flux ratio between the two components.  We first developed a model
of the LBQS quasar composite \ion{C}{4} line.  This line is somewhat
broad and slightly blueshifted, and could be fit well with two
Gaussians, one with a width of $9800\rm \, km\, s^{-1}$ and larger
blue offset $-1520\rm \, km\, s^{-1}$, and the other with a width of
$3750\rm \, km\, s^{-1}$ and smaller blue offset of $-285\rm \, km\,
s^{-1}$.  Our model for SDSS~J0850$+$4451 consisted of two Gaussians
with the same offsets 
and widths as the LBQS \ion{C}{4} model, but variable flux ratio.  We
used this emission line profile model to model \ion{C}{4}, \ion{Si}{4},
\ion{N}{5}, Ly$\alpha$, and \ion{Si}{2}$\lambda 1264$, constraining
the flux ratio between the two components to be equal for the
doublets. The underlying continuum 
could be modeled adequately using a broken power law, but a
4th-order polynomial provided a better fit.  The best-fitting model
and data are shown in Fig.~\ref{fig3}.  The statistical uncertainty in the
spectrum is also shown to highlight the fact that the signal-to-noise
ratio in this spectrum is low for the shortest wavelengths
that cover the diagnostically important \ion{P}{5} line.  Henceforth,
this continuum model is referred to as the ``first continuum model''. 

While this model provided a good fit over the whole bandpass, some
ambiguity remains.  Our solution assumes that the prominent feature
near 1212\AA\/ is unabsorbed continuum and line emission.  However,
that feature may be partially absorbed continuum and line emission
instead, since in our first continuum 
model (shown in red in the figure), the contribution by \ion{N}{5} is
larger than the contribution from Ly$\alpha$.  We created another
continuum estimate by first fitting the mean model spectrum  obtained
by \citet{paris11} from a sample of $z\sim 3$  SDSS quasars with the  
model outlined above, and then requiring the ratio of the intensity of
Ly$\alpha$ to \ion{C}{4} in our spectrum to be equal to the one
obtained from the mean $z\sim 3$ spectrum.  
This model left us with a significant contribution from
\ion{N}{5} emission which is well constrained on the red side of the
line.  The \ion{N}{5} emission line could be enhanced in BALQs by
resonance scattering of Ly$\alpha$  \citep{wang10,hamann96}.  This
model is referred to as the ``second continuum model''.   

One of the uncertainties in modeling broad absorption line
quasars is the placement of the continuum.  In SDSS J0850$+$4451,
there is enough visible continuum between absorption lines that the
only region of significant uncertainty is in the vicinity of
Ly$\alpha$, which we account for as discussed above.  We are currently
developing a technique that models the continuum simultaneously with
the absorption lines \citep{marrs17, wagner17, 
  leighly_aas17}. In any case, SDSS~J0850$+$4451 has deep lines, so
continuum placement is relatively less important for this object than
for objects with shallow lines, where there is little contrast between
the absorption lines and the continuum. 

\section{{\it SimBAL}}\label{simbal}

We provide a brief overview of the {\it SimBAL} analysis method.   Our
model column densities are computed using the photoionization code
{\it Cloudy} \citep{ferland13}.  This mature code is well documented
and maintained, and has a large, constantly updated atomic library.
For this paper, we use data produced by version C13.03 of the code.  A
number of input parameters are required.  For the spectral energy
distribution (SED) of the light illuminating the slab of gas, we began with
a relatively soft SED that may be characteristic of quasar-luminosity
($\gtrsim 10^{46}\rm \, erg\, s^{-1}$) 
objects\footnote{The command to   implement this SED is \tt{AGN
    T=200000K, a(ox)=-1.7, a(uv)=-0.5,     a(x)=-0.9}, where $a$ is
  the power law index.}
\citep{hamann13}.  To check for systematic uncertainty 
associated with SED choice, we also use a relatively hard
SED\footnote{The command to implement this SED is \tt{AGN kirk}, or
  equivalently, \tt{AGN 6.00 -1.40 -0.50 -1.0}.} \citep{korista97}
that may be more 
appropriate for Seyfert galaxies ($\lesssim 10^{45}\rm \, erg\,
s^{-1}$).  Our baseline abundances are the 
default solar abundances (see the {\it Cloudy} manual {\it
  Hazy}\footnote{https://www.nublado.org/wiki/StepByStep} for 
references).  To investigate systematic uncertainty associated with
abundances, we also ran a set of simulations with $Z=3 Z_\odot$,
i.e., all the metals have abundances three times the solar value, with
nitrogen enhanced by a factor of $Z^2$, and helium enhanced by a
factor of 1.14 \citep{hamann02}.  { Other potential systematic
  uncertainties and model dependencies  are discussed
  briefly in Appendix \ref{systematic}}.  

The physical conditions of the gas are parameterized using the
dimensionless ionization parameter $\log U$, the gas density $\log
n\rm \, [cm^{-3}]$, and a combination parameter $\log N_H - \log U$
which essentially measures the column density of the gas  
slab with respect to the hydrogen ionization front,  usually
near 23.2 in this parameterization and depending on the spectral
energy distribution.  Depending on the application, our calculational
grid spans $-4.0 < \log U < 2$, $2.8 < \log n < 9$, and $21.5 < \log
N_H - \log U < 23.7$ with grid spacing 0.05, 0.2, and 0.02,
respectively.  { The broad absorption lines in SDSS J0850+4451 have a
  low minimum velocity, and \ion{C}{4} line is deep and nearly
  black. This indicates that the broad line region is covered, and
  therefore the absorber is outside of the broad line region.
  Therefore, a maximum density of $n=10^9\rm \, cm^{-3}$ is a
  reasonable choice.}  The column densities $N(ion)$ of 179 ground
and excited state atoms and ions were  extracted from the output.
These results were combined with our current line list which includes
6267 transitions.  The result is a large file that is used as input to
{\it   SimBAL}, and from which we synthesize the spectra.    

The opacity as a function of velocity can, in principle, have any
functional form, or be specified by a template.  For a single Gaussian
opacity 
profile, there are six parameters required for the model.  These
include the three gas parameters discussed above ($\log U$, $\log n$, $\log
N_H-\log U$).  Also required are two parameters to describe the
kinematics: the velocity offset $v_{\rm off}$ and the velocity width
$v_{\sigma}$, both specified in $\rm km\, s^{-1}$, and one parameter
to describe the covering fraction of the gas.  While the partial
covering can be parameterized in many ways \citep[e.g.,][]{sabra01},
we choose power-law partial covering, where $\tau = \tau_{max}
x^{a}$.  In this parameterization, $\tau$ is the integrated opacity of
the line, while $\tau_{max}$ is proportional to $\lambda f_{ik}
N(ion)$, where $\lambda$ is the wavelength of the line, $f_{ik}$ is
the oscillator strength, $N(ion)$ is the ionic column density
\citep[e.g.,][]{ss91}, and $x \in (0,1)$ represents the projection of
the two-dimensional continuum source onto a normalized one dimension.
The exponent on $x$ in the form $\log a$ is then the parameter that is
modeled in the MCMC. The power-law partial covering model { has 
been explored by  \citet{dekool02c, sabra01, arav05}, and sometimes
it is found to provide a better fit than the step-function partial
covering model \citep{dekool02c, arav05}.  We choose this
parameterization for the practical reason that because we build the
synthetic absorption spectrum line by line, we need a parameterization
that is mathematically commutative.  The more commonly used step
function covering fraction parameterization is not mathematically
commutative. In addition, the power law parameterization naturally
explains the observation that high-opacity lines are observed to have
a larger covering fraction than low-opacity lines
\citep[e.g.,][]{hamann01}.  

In the power-law covering fraction parameterization, the
fraction of the source covered, or alternatively, the residual
intensity, depends on the total opacity of the line.  Therefore, in a
photoionized gas, where the opacity of different lines can differ
dramatically, so can the residual intensity.  The width of the line is
important, because  one observes $d\tau/dv$, and therefore a wide line
will be less optically thick than a narrow line for the same total
ionic column density.  For reference, the average optical depth of a line in this
parameterization is $\bar{\tau}=\int_0^1 \tau_{max} x^a =
\tau_{max}/(1+a)$.     \cite{arav05} pointed out that for
$\tau(x)=\tau_{max}x^a$, the x-value for which $\tau(x)$ becomes
greater than 0.5 provides a good  estimate of the residual intensity
in the line.  }

\begin{figure*}[!t]
\epsscale{1.0}
\begin{center}
\includegraphics[width=7.0truein]{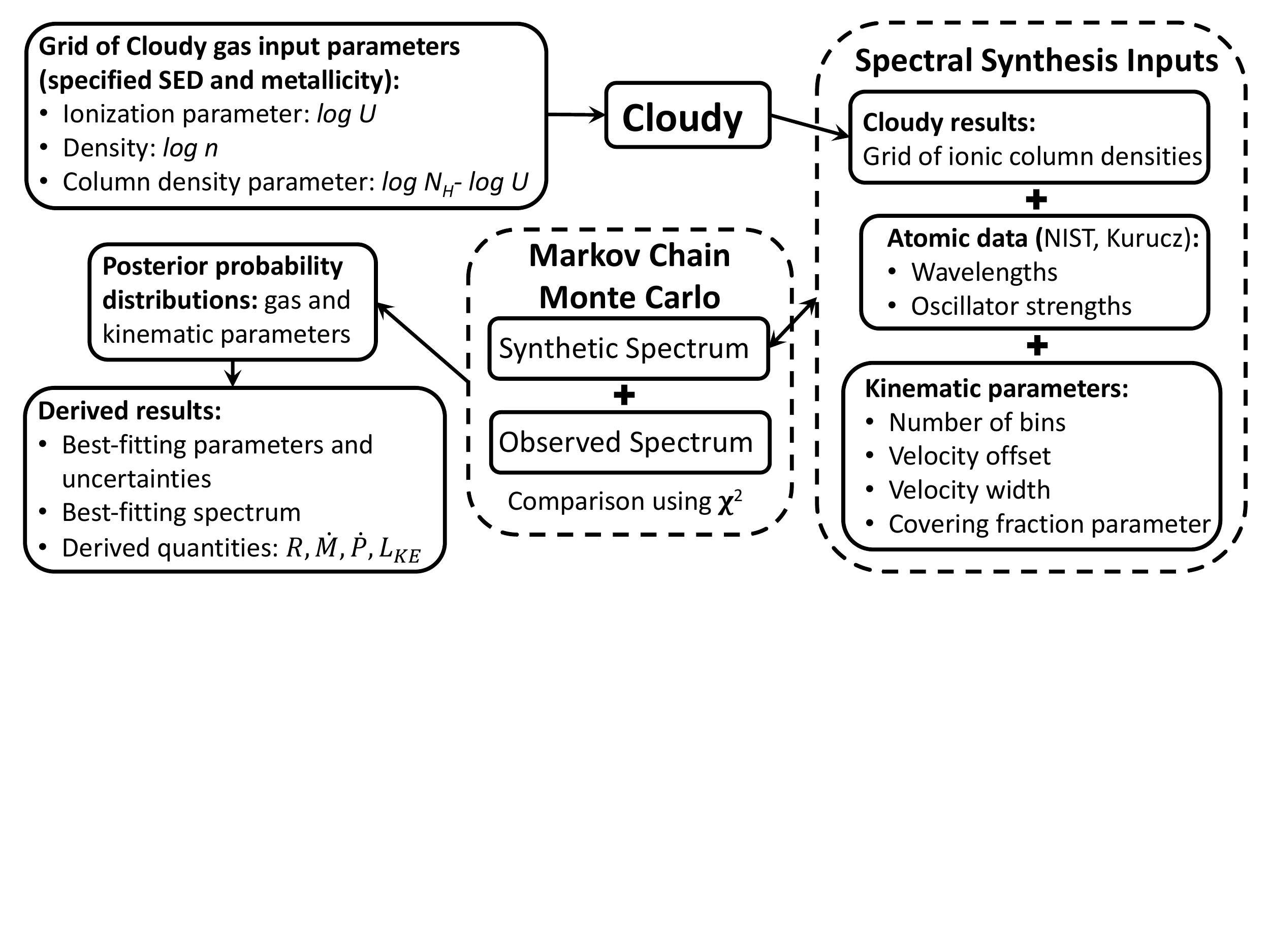}
\caption{A flow chart illustrating the {\it SimBAL} analysis
  procedure.  {\it Cloudy} models are calculated in advance on an
  array of   ionization   parameter ($\log U$), density ($\log n$),
  and a   combination column   density  parameter ($\log N_H-\log U$)
  to yield   a grid of ionic column   densities.  Combined with atomic
  data and   kinematic parameters  including the velocity offset,
  velocity width,   and covering fraction parameter, synthetic spectra
  are generated in   real time.    Using a Markov Chain Monte Carlo
  solver, the 
  synthetic spectra   are compared with the observed  spectrum using
  a likelihood based   on $\chi^2$.  The output is a chain of values
  mapping the posterior    probability distribution of the modeled
  parameters, from which   results such as the best-fitting spectrum
  can be   constructed.   \label{flowchart}}  
\end{center}
\end{figure*}

Once the choice of input {\it Cloudy} matrix and opacity model has
been made, the spectral synthesis modeling can be done.  The synthetic
spectrum is created during the simulations on rest frame wavelengths
of the spectrum to be modeled.  The synthetic spectrum is compared with the
observed spectrum using a likelihood based on  $\chi^2$. Markov Chain
Monte Carlo allows efficient  exploration of parameter space. We use
the {\tt emcee} code\footnote{http://dan.iel.fm/emcee/current/}
\citep{emcee}, which uses the \citet{gw10} affine invariant MCMC
sampler.  This method has the advantage that it can efficiently build
up a posterior probability  distribution even in the face of highly
correlated parameters.  Our model requires specification of a prior
whose minimal constraints keep the model parameters within the
computational boundaries.  We used flat priors for the physical
parameters of the gas, to ensure that the   solution stays within the
computed {\it  Cloudy} model grid.  We used Gaussian priors for the
line offsets and widths, based on inspection, and we checked that the
posteriors of these 
parameters were always narrower than the priors.    The code produces
a chain of values, which, after a period of burnin, can be used to
construct the posterior probability distributions of the free
parameters, or properties derived from them (e.g., the best-fitting
model spectrum, or derived quantities such as the mass accretion
rate).  For this application, we typically do 20,000 simulations using
300 walkers on $\sim 25$ double-threaded cores. A flow
chart of the procedure is shown in Fig.~\ref{flowchart}.  

\section{Absorption-Line Modeling}\label{absorption_modeling}

\subsection{Single Gaussian Opacity Profile}\label{single}

\begin{figure*}[!t]
\epsscale{1.0}
\begin{center}
\includegraphics[width=5.5truein]{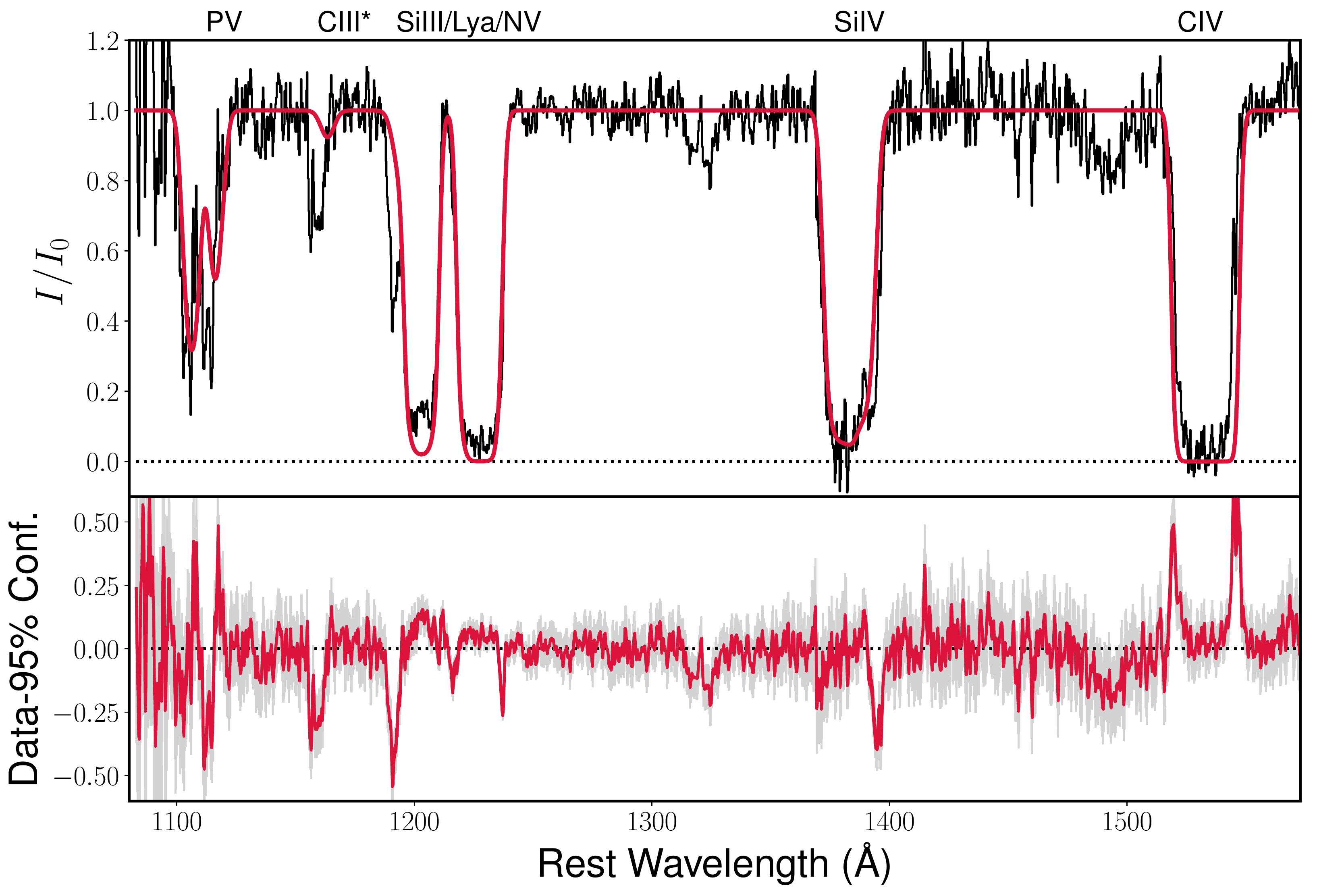}
\caption{The results from a single Gaussian profile model fit to the {\it
    HST} spectrum.   In the top panel, the  median synthetic spectrum
  (crimson) is overlaid on the continuum-normalized spectrum (black).
  The   lower panel shows the spectrum minus the median 
  model and errors in gray, and  the filled region between the
  spectrum and plus   and minus the 95\%   confidence synthetic
  spectra  in crimson, respectively. Significant negative and positive
  residuals are observed,   indicating that the spectrum cannot be
  adequately modeled using a single   Gaussian opacity
  profile.\label{fig5}}  
\end{center}
\end{figure*}

In this section, we present the {\it SimBAL} models of the {\it HST}
spectrum.  To demonstrate the need for a complex model, we began by using a
single Gaussian opacity profile \citep[e.g.,][]{borguet12,
  moravec17}, and performed the MCMC modeling as outlined above.  From
the results we constructed the  median and 95\% confidence
spectra. These are shown in Fig.~\ref{fig5}.   

The figure shows that while the character of the bulk of the
absorption is captured by the single Gaussian model, the absorption is
not modeled well in detail.   Neither the \ion{C}{3}*$\lambda 1175$
feature observed near 1160\AA\/ nor the \ion{Si}{3}$\lambda 1206$
observed near 1190\AA\/ is  modeled well.   In 
addition, the \ion{C}{4} absorption line appears too broad compared
with e.g., \ion{Si}{4}; this occurred because the \ion{C}{4} line is
 optically thick, resulting in significant opacity in the wings of
the lines, while the \ion{Si}{4} is less saturated. 

{ We compute reduced $\chi^2$ for our models as follows.  We
  desire to determine the goodness of fit of the absorption model.  We
  model the continuum ahead of time, and therefore the regions of the
  model spectrum that have the value of 1 cannot contribute to
  determining the goodness of fit.  So we exclude those regions from
  our computation of $\chi^2$.  The reduced $\chi^2$ for this model is
  2.76. Note that we distinguish these measurements of $\chi^2$ from
  the likelihood used in the MCMC (above).  There, the lack of the
  line is important for constraining models, so we use the full
  wavelength range.  }

\subsection{Accordion Models}

From the shape of the absorption lines and residuals shown in
Fig.~\ref{fig5}, it is clear that the spectrum can only be explained
adequately with 
multiple  absorbing components.  An obvious next choice is a model
consisting of multiple Gaussians.  A problem with multiple Gaussians with all
parameters free is that they can move enough in the spectral fitting
to mix with one another in the MCMC.  We found that a constrained
multicomponent model is more robust in the spectral fitting, and has
the added benefit of yielding information about the physical
conditions as a function of velocity.  

We call the constrained model that we used an ``accordion model''.
The model is characterized by multiple components that we call
``bins'' to emphasize that these are not to be considered separate
velocity components in the outflow, but rather a parameterization of
the opacity as a function of velocity.  Each bin has the same width
and separation from its neighbors in velocity space.   The fitting
parameters are then the maximum velocity offset (i.e., the central
velocity of the shortest-wavelength bin), the 
velocity width of the bins, the separation of the bins (3 parameters),
and, most generally, $\log U$, $\log n$, and $\log N_H-\log U$ for
each bin ($3x$ the number of bins).  We tried two
versions of the accordion model: one used a Gaussian bin, and the
other used a tophat bin.  The tophat accordion model had one fewer
parameter than the Gaussian accordion  model because the width of a
bin is equal to the separation. 

We rejected the Gaussian accordion model for the following reasons.
In the tophat accordion model, the opacity of each bin is
independent of its neighbors; this is by construction. To put it
another way, there is no blending of adjacent bins.  Therefore, in a
tophat accordion model, for a feature comprised of a single 
transition, a plot of the individual bins will touch the minimum flux
points of the synthesized spectrum  and therefore follow the outline
of the feature smoothly 
(except for the stepped approximation).  In contrast, the opacity of
bins are not independent in the Gaussian accordion model, because
the wings of adjacent Gaussian bins must overlap, even if the
individual bins are narrow.
So in a corresponding plot of individual bins for the Gaussian
accordion model, the minimum values of the individual bins will not touch
the minimum flux level of the synthetic model spectrum; they all
appear optically thinner (or 
appear to have a lower covering fraction) than the feature does
\citep[e.g.,][Fig.\ 11]{rupke02}.  We suggest that this lack of
independence results  in unreliable inferences about the wind
properties for the main features of the absorption, although there is
evidence that the Gaussian accordion model can pick up low covering
fraction features with different properties (e.g., $\log U$ or $\log
N_H - \log U$) but the same velocities as
the main features.  We therefore reject the Gaussian accordion model 
and only consider the tophat accordion model for the remainder of this
paper. 

We first present a tophat accordion model with 11 bins.  In this
model, the ionization parameter, column 
density parameter $\log N_H - \log U$, and covering fraction power-law
index $\log a$
were allowed to vary freely for each bin.  The density was allowed
to have two values, depending on the bin, with the reasoning for
that  choice as follows.  \ion{C}{3}*$\lambda 1175$ is a transition
from a metastable state (the decay from that state produces the 
\ion{C}{3}]$\lambda 1909$ emission line).  The level is complex, with 
$J=0$, $J=1$, $J=2$ states.  These states have different transition
probabilities and therefore have different critical densities
\citep[e.g.,][]{gabel05}. These properties make the
\ion{C}{3}*$\lambda 1175$ is a 
density-sensitive line \citep[see][for   a discussion of the utility
  of density-sensitive lines for   determining absorber
  distances]{arav13}, but rarely used because the transitions are
close together and cannot be deblended except for the narrowest of
absorption lines \citep{gabel05, borguet12}.  The spectral synthesis
approach does not require deblending, so we can use this line to
determine the density of the portion of the outflow (i.e., velocity
range) represented by this line.  For 11 bins, that range was the 4th,
5th, and 6th bins from the left (i.e., spanning $-3200$ to $-4450\rm
\, km\, s^{-1}$).  As shown below, this feature is characterized by a
larger value of $\log N_H - \log U$, and we refer to this enhanced
column density region in the outflow as the
``concentration'' henceforth.   The densities of these three
high-column-density bins were constrained to have the same value, and
the densities of the remaining seven were constrained to have a
different value.  Thus, two values of density were fit.

\begin{figure*}[!t]
\epsscale{1.0}
\begin{center}
\includegraphics[width=5.5truein]{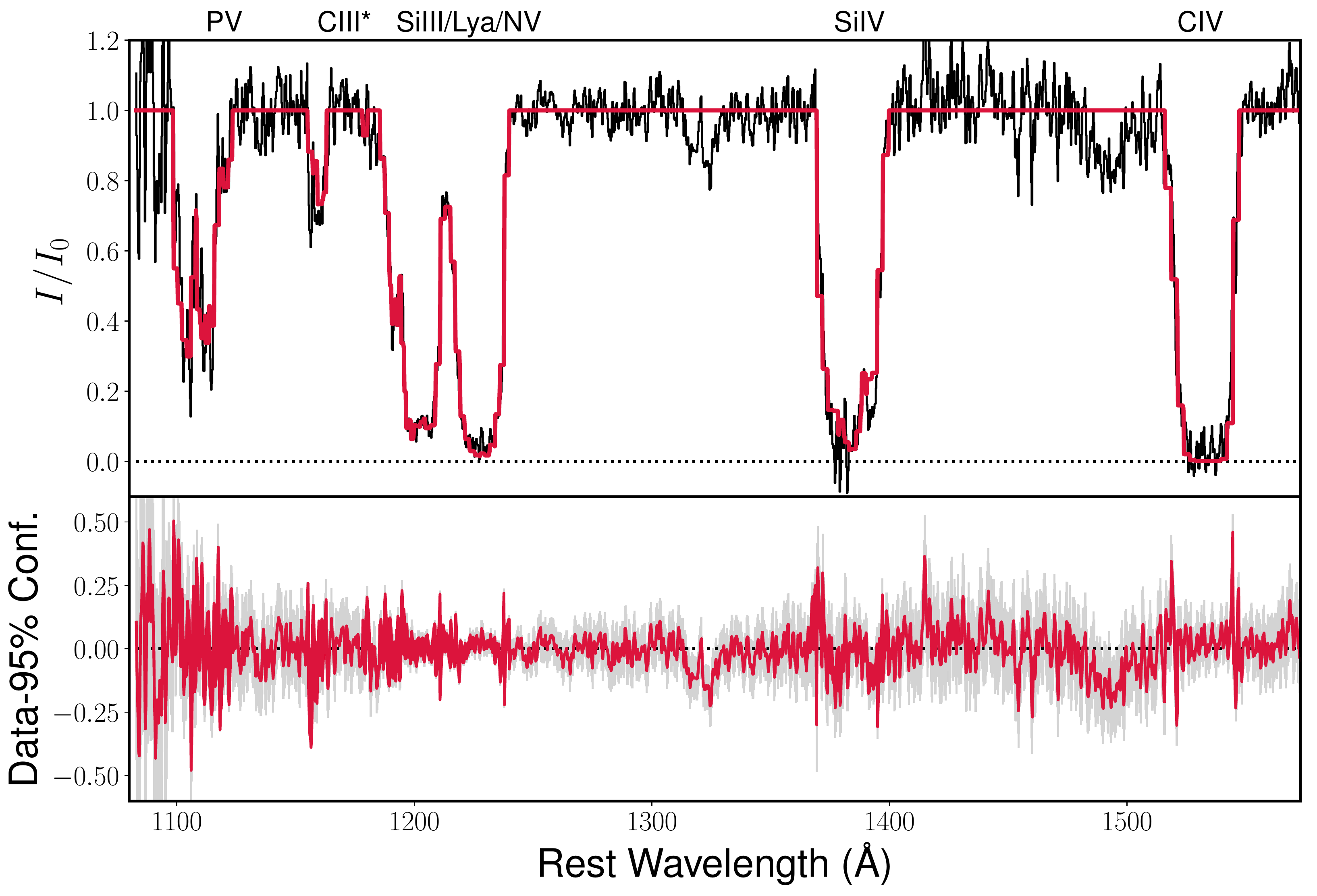}
\caption{The results from an 11-bin tophat accordion model.  In the top panel, the  median synthetic spectrum
  (crimson) is overlaid on the continuum-normalized spectrum (black).
 The   lower panel shows the spectrum minus the median
  model and errors in gray, and  the filled region between the
  spectrum and plus   and minus the 95\%   confidence synthetic
  spectra in crimson, respectively.  Overall, the fit is good.
  Negative residuals near 1160, 1320, and 1490\AA\/ are plausibly an
  intervening Ly$\alpha$   forest line, \ion{C}{2} characterizing
  slightly higher values of $\log N_H-\log U$ than are represented
  here, and \ion{C}{4} from a   low-opacity high-velocity outflow,
  respectively.\label{fig6}}   
\end{center}
\end{figure*}

\begin{figure*}[!t]
\epsscale{1.0}
\begin{center}
\includegraphics[width=4.5truein]{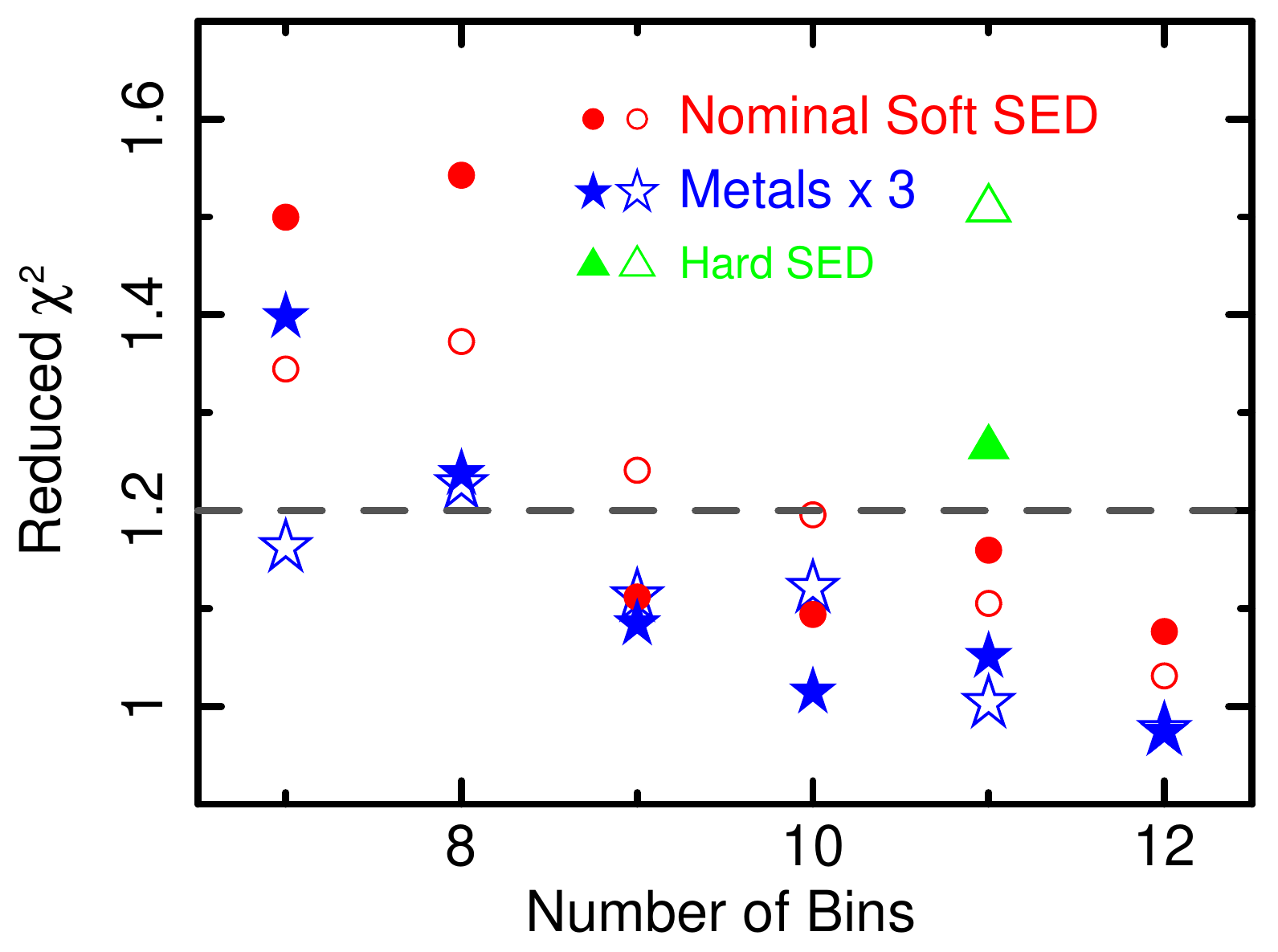}
\caption{Reduced $\chi^2$ for the 26 tophat accordion models presented
  in this paper, evaluated between 1080\AA\/ and 1573\AA\/ { as
  described in \S\ref{single}.} In each
  case, the open symbols show the first continuum model, while the
  solid  symbols show the second continuum model (see
  \S\ref{cont_model}).  { Formally, values of reduced $\chi^2$ greater
  than $\sim 1.2$ (dashed line) are unacceptable at 99\% confidence.
  We find that   the larger the number of bins, the better the fit,
  the   $Z=3Z_\odot$  metallicity provides a   slightly   better fit
  than   solar metallicity, and the hard SED   yields a relatively
  poorer   fit than the soft SED.}\label{fig4}}  
\end{center}
\end{figure*}

The model and the difference between the data and the median spectrum
for the second continuum model is shown in Fig.~\ref{fig6}; the
results for the first continuum model are very similar.   The fits are very
good overall.  The reduced $\chi^2$ for this model and others
discussed below are shown in Fig.~\ref{fig4}.  As discussed above, in
order to ascertain goodness of fit of the absorption model, we
compute the reduced $\chi^2$ over regions of the spectrum where the
model absorption optical depth is greater than zero.  For the typical number
of degrees of freedom (around 500), a value of reduced $\chi^2$
greater than 1.2 is excluded at a 99\% confidence level.  Generally,
the second continuum model produces a slightly better fit than the
first continuum model.

There are several notable residuals.  Just left of the
\ion{C}{3}* line at 1160 \AA\/ is a  sharp line-like feature, which is 
plausibly a Ly$\alpha$ forest line, although some models fit this
feature and some do not. Around 1320 \AA\/ are a pair of
features that are consistent with 
\ion{C}{2}$\lambda 1335$ at $v=-4120\rm \, km\, s^{-1}$ and
$v=-2460\rm \, km\, s^{-1}$, i.e., within the velocity range of the
main feature.  \ion{C}{2} is a low ionization line that appears at
slightly larger values of $\log N_H-\log U$ that are represented in
this object (see Fig.~\ref{fig9b})    Finally, there is a broad
feature centered around $\sim 
1490$ \AA\/ that probably originates in high-velocity
gas, centered around $\sim -11200 \pm 1300 \rm \, km\, s^{-1}$
(1$\sigma$), that is of low optical depth so that only 
C$^{+3}$ presents significant opacity.  Ly$\alpha$ absorption from
this feature is predicted to lie between 1166 and 1176\AA\/, and there
is no absorption observed in that band. Absorption from  \ion{N}{5}
could plausibly be blended with the observed \ion{Si}{3}$\lambda 1207$
/ Ly$\alpha$ feature, but it is likely to be quite shallow if present.
We ignore these features  henceforth.   

\begin{figure*}[!t]
\epsscale{1.0}
\begin{center}
\includegraphics[width=5.5truein]{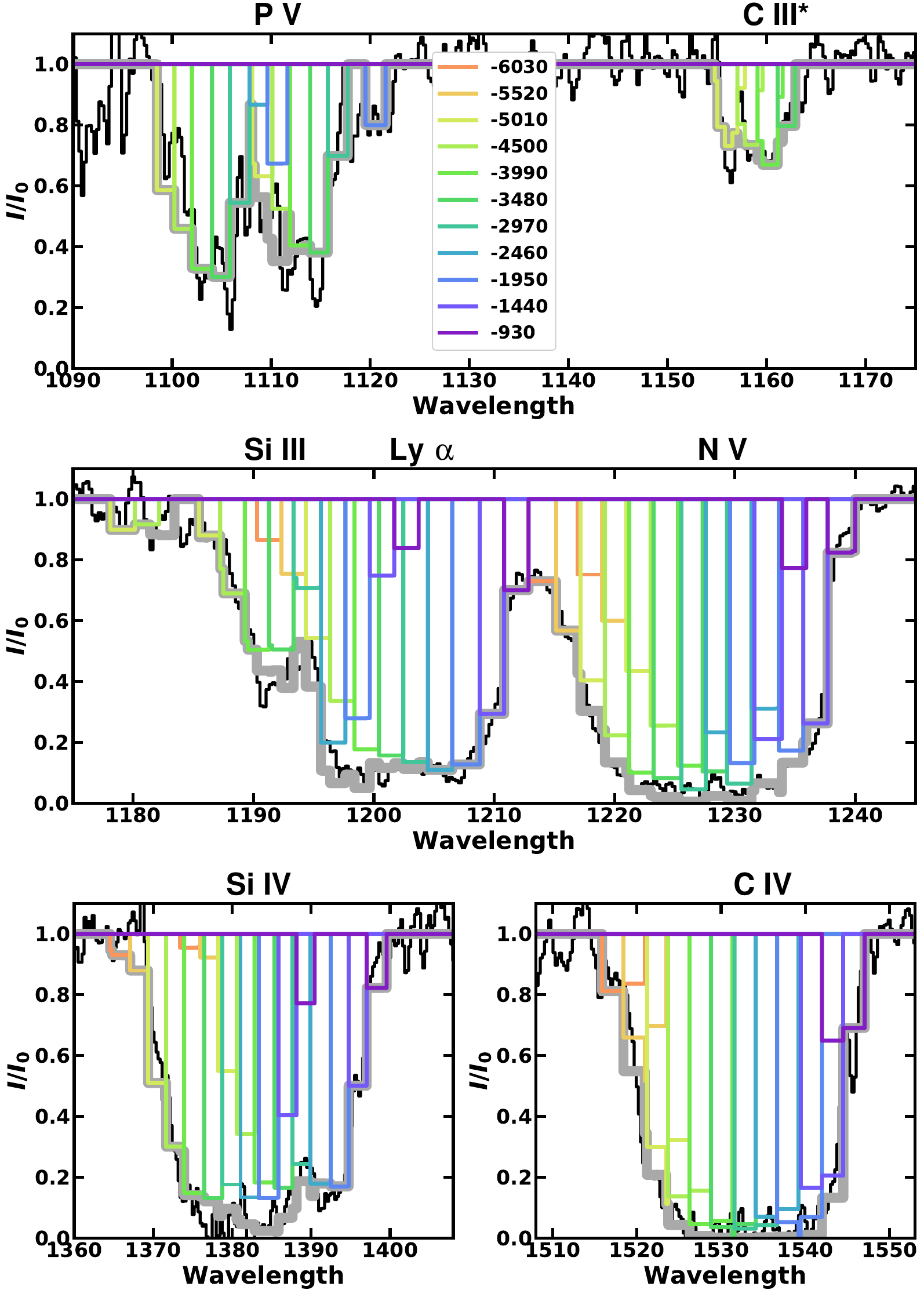}
\caption{The results from an 11-bin tophat accordion model for the
  second continuum model and enhanced metallicity ($Z=3 Z_\odot$); see
  \S~\ref{metallicity}.  Each of the bins is shown by a
  different color, and the full line profile is shown in a thick light
  gray line   overlaid on the continuum-normalized data (black). The principal
  lines   contributing to the opacity are identified over the
  features.  The   opacity for each member of the resonance doublets
  \ion{P}{5},   \ion{N}{4},  \ion{Si}{4}, and \ion{C}{4} can be seen.
  The    limited extent of   the density-sensitive \ion{C}{3}* line
  can be seen; only 
  four of   the eleven bins, with center velocities between $-5010\rm
  \,   km\, s^{-1}$ and $-3480\rm  \, km\, s^{-1}$, contribute this
  this   line.    \label{fig6b}}  
\end{center}
\end{figure*}

Fig.~\ref{fig6b} displays the individual velocity bins for the case of
the second continuum model and $Z=3 Z_\odot$ 
abundances (see \S~\ref{metallicity}).  The absorption as a function of
velocity is shown, with a different color for each velocity bin.
Individual members of the doublets can be distinguished in \ion{P}{5},
\ion{N}{5}, \ion{Si}{4}, and \ion{C}{4}. For this model, the width /
separation of the bins is rather tightly constrained in order that
the opacity for Ly$\alpha$ and \ion{N}{5} meet to form the
low-absorption region near 1212\AA\/.  It is also notable that the
best-fitting bin width is such that the \ion{C}{4} doublet lines are
in adjacent bins, while the \ion{Si}{4} doublet 
lines are separated by four bins.  This result makes sense since the doublet
separation for \ion{C}{4} is $492 \rm \, km\, s^{-1}$, while the
doublet separation  \ion{Si}{4} is $1932\rm \,  km\, s^{-1}$, i.e., 4
times larger.

What is the effect of the number of bins?  To investigate this
point, we considered models with 7, 8, 9, 10, 11, and 12
bins.  As above, the ionization parameter, $\log N_H -\log U$, and
covering fraction $\log a$ were allowed to vary freely in each bin,
and two densities were used.  The number of bins spanning the
concentration varied, generally increasing as the number of bins was  
increased and as each bin  became narrower.  The
quality of the fit improved as the number of bins was increased, 
likely  due to the greater flexibility of the model; the reduced
$\chi^2$ in the 1080\AA\/ to 1573\AA\/ region as a function of the
number of bins is shown in Fig.~\ref{fig4}. {  As noted above, for
  the typical number of degrees of freedom, a value of reduced
  $\chi^2$ greater than 1.2 is excluded at 99\% confidence level,
  suggesting that the 7- and 8-bin cases do not provide sufficient
  flexibility to model velocity dependence of the absorption
  profiles.}

\begin{figure*}[!t]
\epsscale{1.0}
\begin{center}
\includegraphics[width=7.0truein]{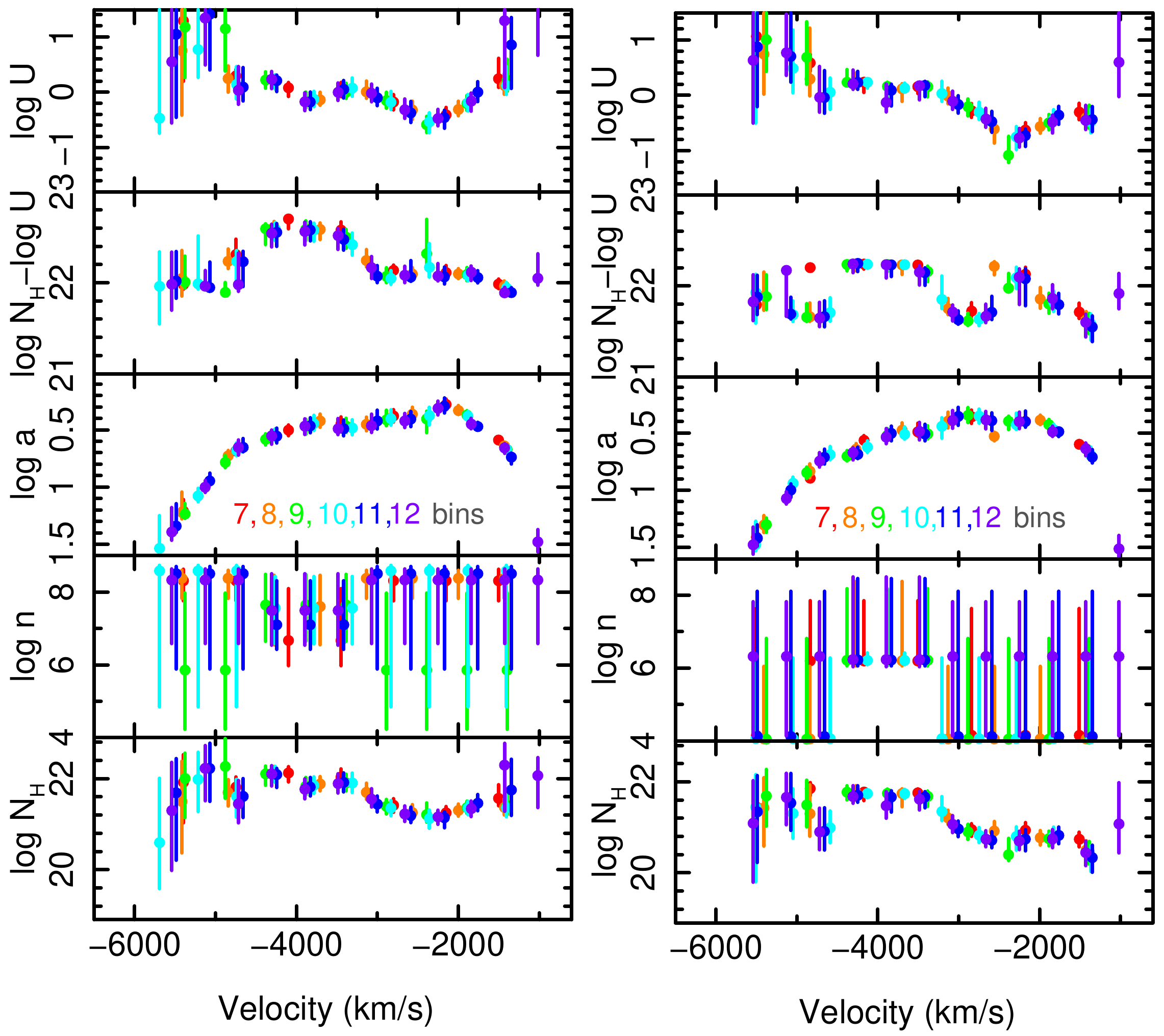}
\caption{The spectral fitting results as a function of velocity
  offset for a range of number of bins spanning the absorption
  profile.  The Maximum Amplitude Probability (MAP) values and 95\%
  confidence regions from the posterior probability distributions are
  shown.  
 The left (right) panel shows the results for solar ($Z=3 Z\odot$)
 metallicity, both for the soft SED and the first continuum model. In
 each of the 12 models (six bin  values and two   metallicities), the
 ionization parameter $\log  U$, the column density parameter $\log
 N_H-\log U$, and the  covering fraction parameter $\log a$ were allowed to vary
 independently in each bin, while the density was allowed to take on
 one of two   values depending on whether or not a bin was
 represented in   the   density-sensitive line \ion{C}{3}*$\lambda
 1175$ which spans the  enhancement in $\log N_H-\log U$ between
 $-4500$ and $-3500\rm\, km\, s^{-1}$ (identified in the text as the
 concentration).  As   expected,   the density is constrained only
 over the   ranges of   velocities where   the \ion{C}{3}*$\lambda
 1175$ line is   present.   The column density   parameter varies
 significantly as a   function   of velocity, being   about a factor
 of three larger between    $-4400$ and $-3200\rm \,   km\,s^{-1}$,
 and the covering fraction   decreases strongly as a   function of
 velocity.    \label{fig7}}   
\end{center}
\end{figure*}

\begin{figure*}[!t]
\epsscale{1.0}
\begin{center}
\includegraphics[width=7.0truein]{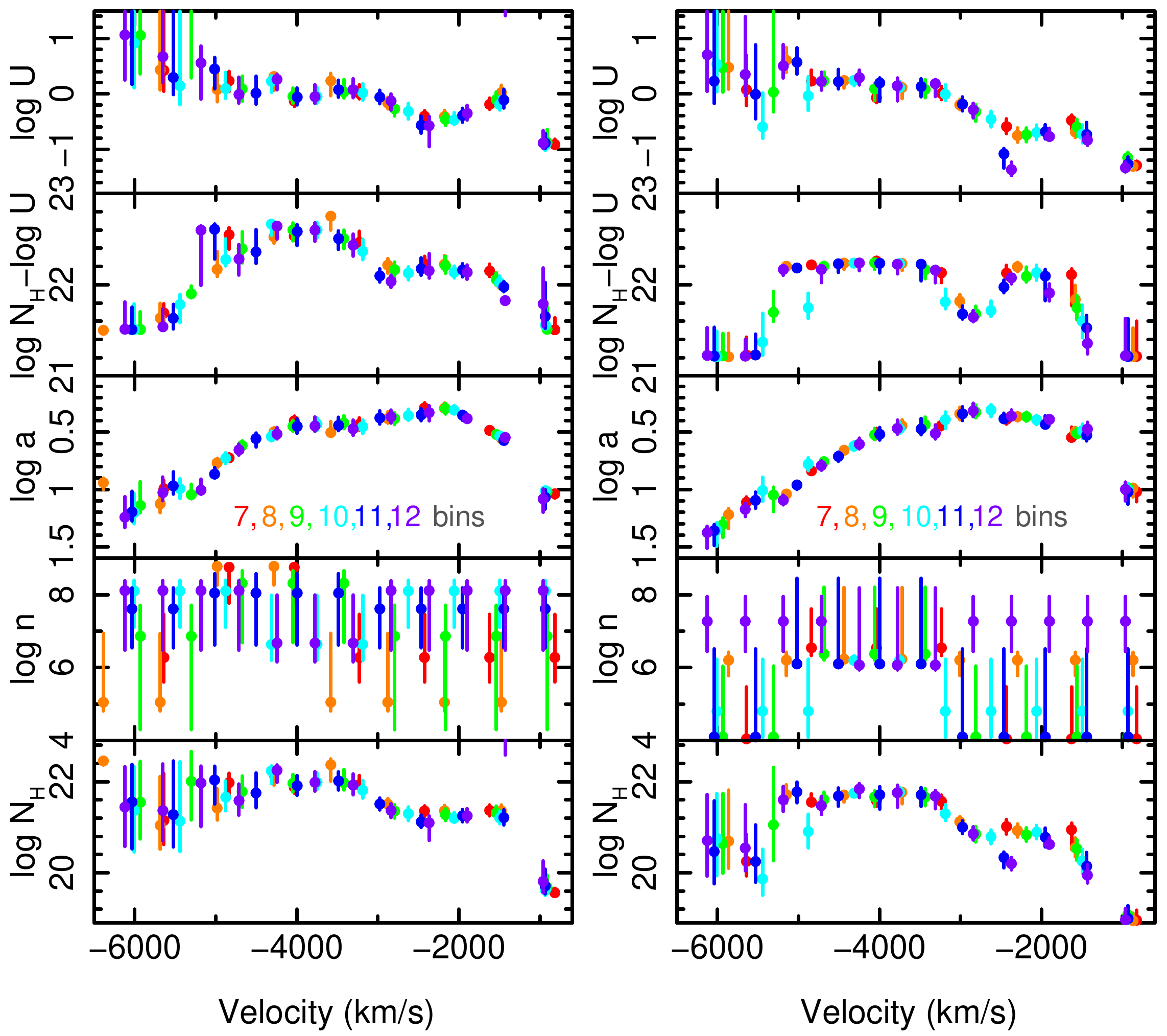}
\caption{Same as Fig.~\ref{fig7}, for the second continuum
  model.  The principal differences lie at the
  lowest velocities and the highest velocities.  The low-velocity end
  of the Ly$\alpha$ absorption and the high-velocity end of the
  \ion{N}{5} absorption abut at the high point  between these two
  features (see Fig.~\ref{fig6b}).  This requirement has the effect of
  strongly   constraining allowed values of velocity for the bins.     \label{fig8}}     
\end{center}
\end{figure*}

Fig.~\ref{fig7} shows the parameter constraints obtained from the posterior
probability distributions (i.e., the MCMC chain excluding burnin) for
all 6 number-of-bin combinations for the first continuum model,
and Fig.~\ref{fig8} shows the same result for the second continuum
model.  In each case, the best-fitting point is the Maximum Amplitude
Probability (MAP) of the posterior probability distribution, and the
error bars are obtained 
from the 4.6\%  and 95.4\% (i.e., $2\sigma$) points on the cumulative
distribution.  The results are shown for the four fitted parameters
($\log U$, $\log N_H - \log U$, covering fraction index $\log a$, and
$\log n$). { We also graph the column density $\log N_H$, weighted
  by the covering fraction, i.e., the amount of gas in the outflow,
  which is computed from the sum of $\log N_H-\log U$, $\log U$,
and  $\log (1.0/(1.0+a))$ at each point in the chain.}

Several trends are apparent. The parameters are better constrained in
the middle of the feature, where the absorption is deepest, compared
with the wings. The ionization parameter shows a slight increase
towards higher velocities and a dip near $-2500 \rm \, km\,
s^{-1}$. The column density parameter, 
$\log N_H - \log U$ is clearly higher by a factor of 3  in the
vicinity of the concentration.   The  density is
poorly constrained, even when only two values are used,  but it is
better constrained (by \ion{C}{3}*) in the vicinity of the
concentration  as compared with the values outside the
concentration. 

The covering fraction  parameter varies
strongly with velocity, with a lower covering fraction  at higher
velocities.  In the vicinity of the concentration, the  covering
fraction $\log a=0.5$, and at higher velocities, the covering fraction
$\log a  \sim 1$.  { The covering fraction is the most strongly
  variable parameter, indicating that it is important in determining
  the shapes of the troughs.  The strong dependence on covering
  fraction is   consistent with the behavior of other outflows, where
  the apparent optical depths of absorption lines are found to be controlled
  principally by covering fraction, rather than the ionic column
  density as one might expect \citep[e.g.,][and references
    therein]{arav04}. Note that the covering fraction variations do
  not mirror the photoionization properties of the gas; for example,
  the covering fraction maximum, located around $-2500\rm \, km\,
  s^{-1}$, is not at the same velocity as the 
  $\log N_H-\log U$ maximum, located between $-3200$ to $-4450\rm
\, km\, s^{-1}$, i.e., the concentration. }

The covering-fraction weighted column density is roughly constant for
velocities faster than  $-3300\rm \, km\, s^{-1}$, with an average
column over that region of $\log N_H=22.0\rm \, [cm^{-2}]$.  This
region includes the concentration, between $-4400$ and $-3300\rm \,
km\, s^{-1}$, which exhibits a distinctly greater value of $\log N_H
-\log U$ (Fig.~\ref{fig6}).  But it also includes, interestingly, the
highest velocity portions of the outflow, which have lower value of
$\log N_H  - \log U$  and a higher value of $\log a$ (corresponding to
a lower covering fraction), which would tend to decrease the inferred
column density.  Those two factors are offset by a higher value of
$\log U$ approaching 1, corresponding to a thicker \ion{H}{2} region,
which offsets the effects of  the lower $\log N_H -\log U$ and higher
$\log a$.  The error bars are large, as the feature is shallow at the
highest velocities, so this result is uncertain.   The column density
decreases at lower velocities, to an approximate constant level for
velocities smaller than $-2800\rm \, km\, s^{-1}$, of $\log N_H=21.4 \,
\rm [cm^{-2}]$.   

The ionization parameter and covering-fraction weighted column
densities are both larger at high velocities and smaller at
low velocities, suggesting a correlation between the ionization 
parameter and column density. 

The results for the second continuum model are shown in Fig.~\ref{fig8}.
Overall, they are qualitatively similar.  The differences principally
originate in the requirement that some opacity, but not too much, be
present on the boundary between the \ion{N}{5} and Ly$\alpha$ lines,
resulting in  rather stringent constraints on the widths of the
bins; this forced the whole feature to be wider for the second
continuum, and to have lower opacity at the highest and lowest
velocities. It is also responsible for the relatively poorer fits for
the 7- and 8-bin models.

\subsubsection{What Drives the Fits?}  

The MCMC results show relatively small uncertainties in the  model
parameters, with the exception of $\log n$, which is in general
poorly constrained.  In addition, some of the fit parameters vary as a
function of velocity.  It is therefore interesting to investigate what
properties of the spectrum result in these well-fitting and relatively
tightly-constrained models.  We did this by isolating the best fit
(median values from the posterior) and then varying one parameter
while leaving the others fixed at the best value.  

\begin{figure*}[!t]
\epsscale{1.0}
\begin{center}
\includegraphics[width=7.0truein]{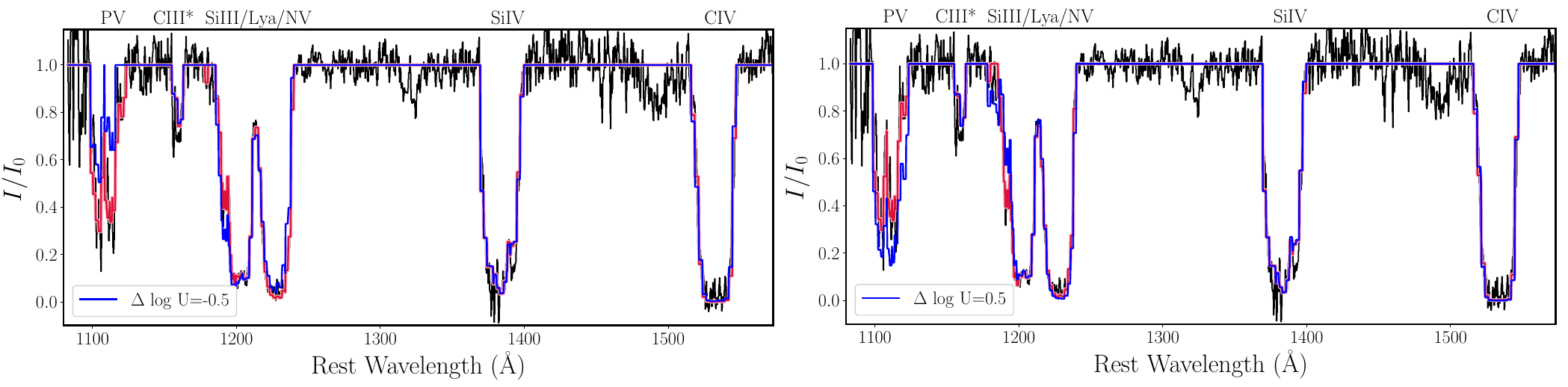}
\caption{Varying parameters away from the best fit demonstrates that
  certain individual lines or groups of lines are responsible for the
  relatively well-constrained parameters obtained.   The results (blue
  line) for  varying the ionization parameter away from the best fit
  (red line) for   the 11-bin tophat accordion model are shown, with
  the negative change in the left panel, and the positive change in
  the right panel.   The figure shows that  \ion{P}{5} is principally
  responsible for   constraining the ionization parameter.  This figure is available as
  an animated movie online. \label{fig9a}}     
\end{center}
\end{figure*}

\begin{figure*}[!t]
\epsscale{1.0}
\begin{center}
\includegraphics[width=7.0truein]{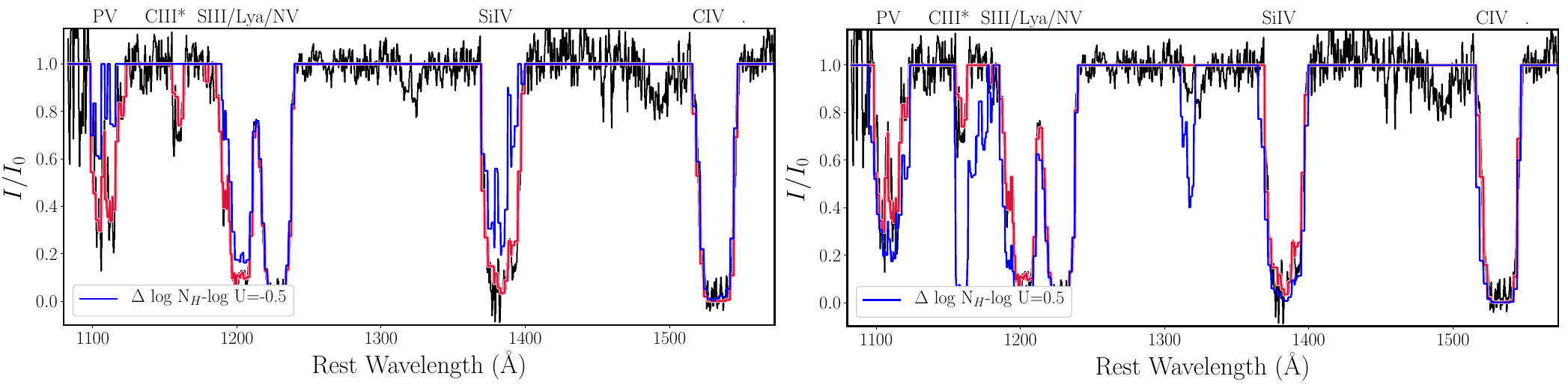}
\caption{The same as Fig.\ ~\ref{fig9a} for $\log N_H -\log U$.  The
   plot shows that correlated variability among several lines,
   especially \ion{C}{3}* and \ion{Si}{4}, as well as lines that are
not observed (such as \ion{C}{2}$\lambda 1335$) are   responsible for
constraining  $\log N_H -\log U$.  This figure is available as an
animated movie   online. \label{fig9b}}    
\end{center}
\end{figure*}

\begin{figure*}[!t]
\epsscale{1.0}
\begin{center}
\includegraphics[width=7.0truein]{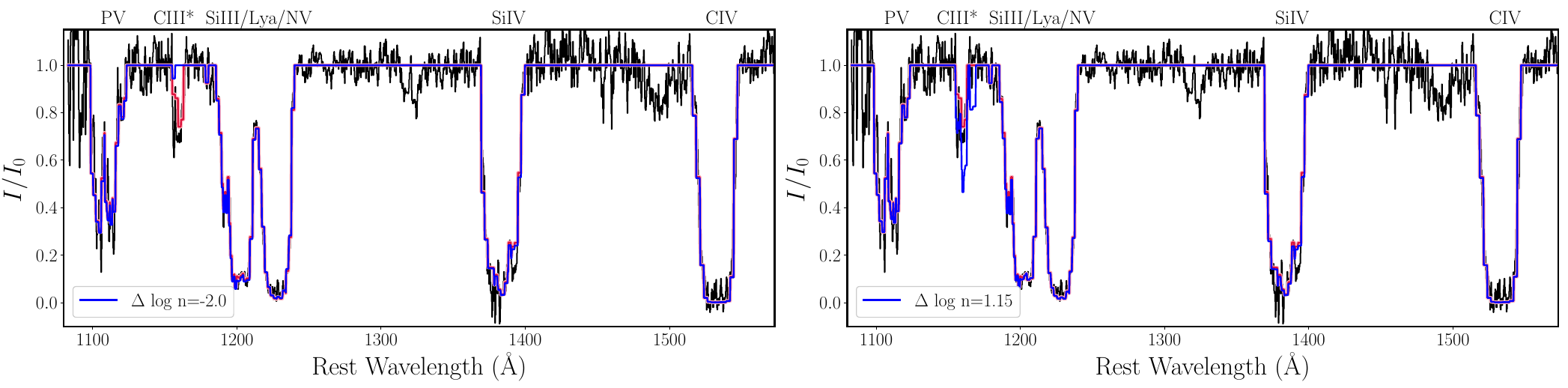}
\caption{The same as Fig.\ ~\ref{fig9a} for covering fraction
  parameter $\log a$.  Note that a larger value of $a$ corresponds to
  a lower covering fraction (see \S~\ref{simbal}). The plot shows that
  varying $\log a$ causes all lines to   vary together.  This figure
  is available as an animated movie 
  online. \label{fig9d}}    
\end{center}
\end{figure*}

\begin{figure*}[!t]
\epsscale{1.0}
\begin{center}
\includegraphics[width=7.0truein]{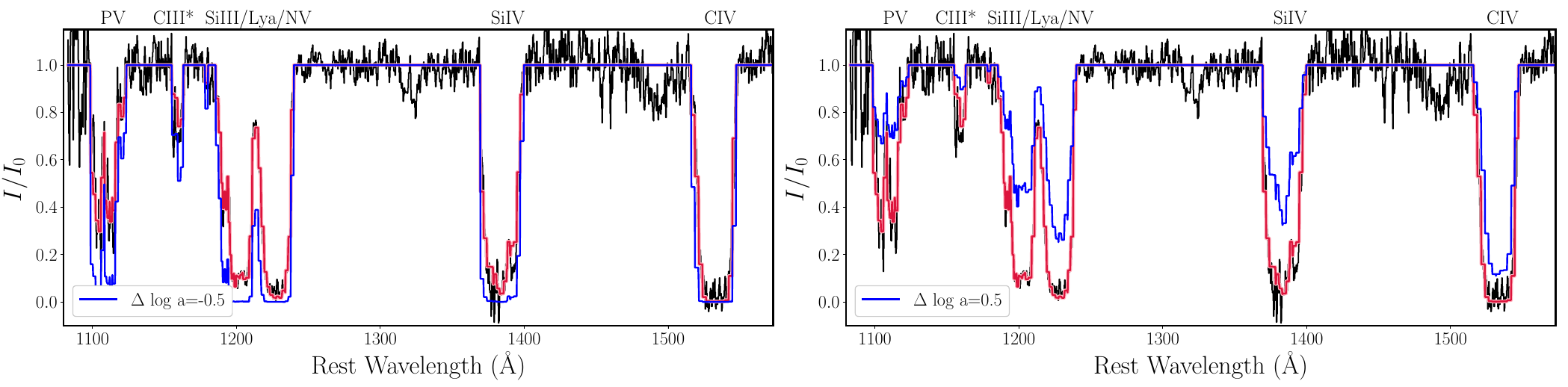}
\caption{The same as Fig.\ ~\ref{fig9a} for the density $\log n$,
  with the exception that the log density is varied to $\Delta \log
  n=-2.0$ and $\Delta \log n=1.15$ to account for the lower
  sensitivity of the simulations to this parameter.  
  Only \ion{C}{3}* is   sensitive to density in this bandpass.  This
  figure is available as   an animated movie   online. \label{fig9c}}     
\end{center}
\end{figure*}

Fig.~\ref{fig9a}--\ref{fig9c} show two still frames from each of the
animations that are available online.  In each figure, the red line
shows the best fitting median model spectrum, while the blue line
shows the model spectrum for minus and plus 0.5 dex from the best fit
for each of the varied parameters (with the exception the plot for
$\log n$, which is less sensitive to variation).

The simulation in Fig.~\ref{fig9a} shows that it is the \ion{P}{5} 
that constrains the ionization parameter in the vicinity of the best
fit.  Conventionally, \ion{P}{5} is thought to be a 
diagnostic of column density, not ionization parameter
\citep[e.g.,][]{hamann98a}.  Since phosphorus is a factor of
765 times lower in abundance than carbon, observation of a
significant \ion{P}{5} line implies a strongly saturated \ion{C}{4}
line and a high column density \citep[e.g.,][]{leighly09, borguet12}.
In addition, P$^{+4}$ and C$^{+3}$ are the dominant ionization state
of their respective elements at nearly the same ionization parameter
of about $\log U=-2$ \citep[e.g.,][Fig.\ 2]{hamann97a}.  { However,
because of the low abundance of phosphorus at $\log U=-2$, there may
not be enough P$^{+4}$ to produce an observable line, even if the
column density is large and extends to the hydrogen ionization front.
Whether or not there is a detectable line depends on the width of the
line, since what is observed in a spectrum is  $d\tau/dv \propto
dN_{ion}/dv$, which means that a small ionic column density can
produce a narrow line that is deep, but a much higher ionic column
density is  required to produce a broad line that is deep.  The thickness of
the Str\"omgren sphere 
increases with $U$, which means that more P$^{+4}$ ions are available
for larger $U$.  Line ratios are approximately constant for a
particular value of $\log N_H-\log U$, as long as the ionization
parameter is not too far from the value where that ionization state is
dominant.  So a larger value of $\log U$ for a given value of $\log
N_H - \log U$ yields a larger column density of all ions, including
enough P$^{+4}$ so that a broad absorption line can be produced.}

The simulations shown in Fig.~\ref{fig9b} reveal that lower values of
the column density parameter $\log N_H-\log U$ are ruled out because
they don't predict sufficient absorption from lower-ionization lines
such as \ion{Si}{3}, \ion{Si}{4}, and \ion{C}{3}*.  This result makes
sense,  since the parameter $\log N_H- \log U$ measures the thickness
of the gas relative to the hydrogen ionization front.  As the gas
becomes thicker and the continuum loses energetic photons, lower
ionization ions start to become more prevalent
\citep[e.g.,][Fig.\ 1]{hamann02}.  Larger values of this parameter are
ruled out by the same ions, which start to produce lines that are
stronger than we see. The \ion{C}{3}* feature increases especially rapidly.

Fig.~\ref{fig9d} shows that the covering fraction parameter
is  relatively tightly constrained.  Below the best-fitting value for
$\log a$, the lines are deeper and some of them appear to be black,
while above the best-fitting value, the lines are quite shallow, even
though the column density held constant in this set of simulations.

Fig.~\ref{fig9c}, showing the change in the model as a function of
density, illustrates what we expect: only \ion{C}{3}*$\lambda 1175$
changes substantially as the density changes, but even then, the
change is quite subtle.  Therefore,  density is not as well
constrained as the other parameters, but it is constrained in the
region of the concentration where there is sufficient \ion{C}{3}*
optical depth.

These figures illustrate the strong complementary physical constraints
on the properties of the gas available over this small bandpass that
can only be practically harvested using synthetic spectral analysis.

\subsubsection{Effect of the Spectral Energy Distribution}\label{sed}

\begin{figure*}[!t]
\epsscale{1.0}
\begin{center}
\includegraphics[width=5.5truein]{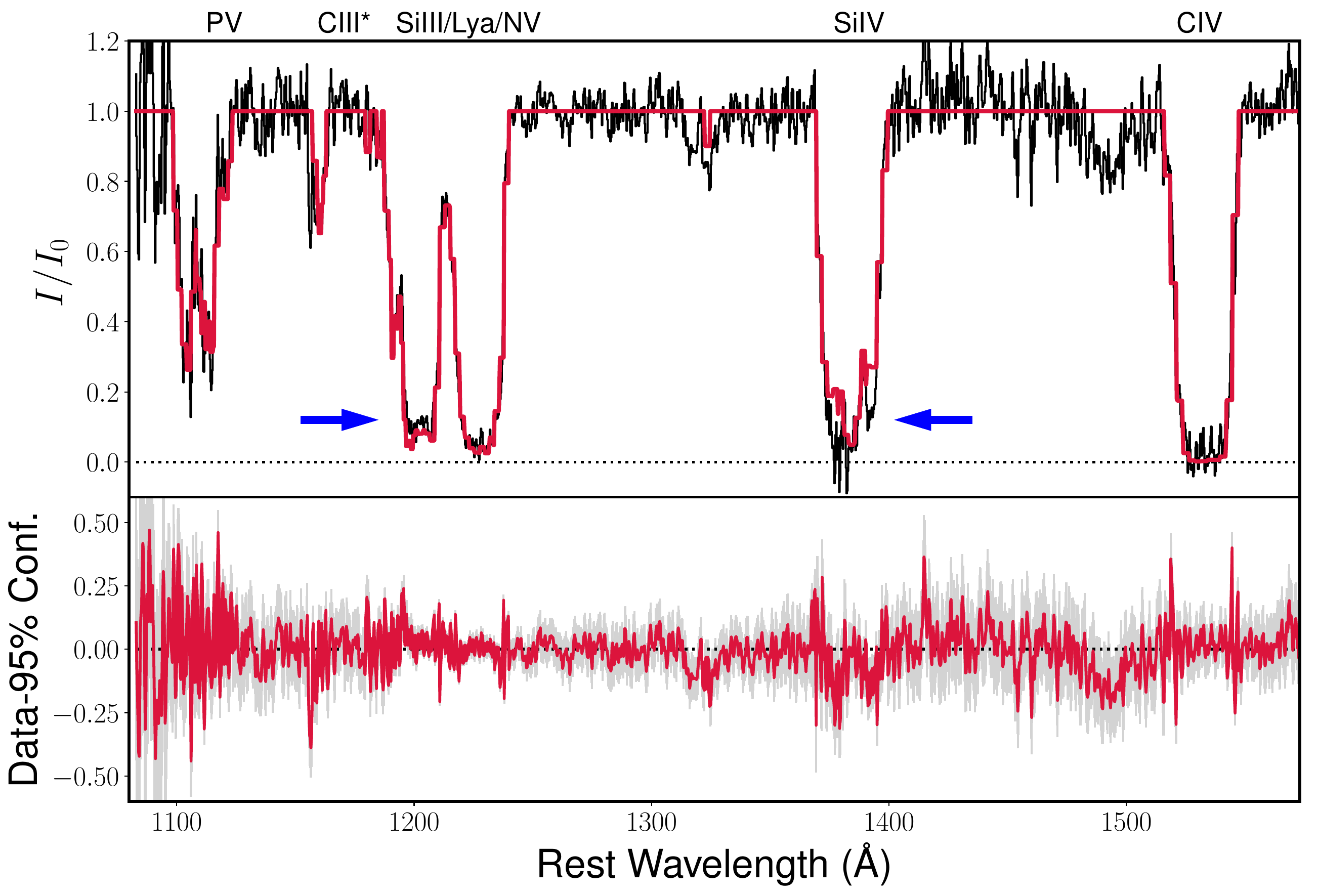}
\caption{The results from an 11-bin tophat accordion model, using
  ionic columns from {\it Cloudy} runs using a hard spectral energy
  distribution \citep{korista97}.  In the top panel, the  median synthetic spectrum
  (crimson) is overlaid on the continuum-normalized spectrum (black).
  The   lower panel shows the spectrum minus the median 
  model and errors in gray, and  the filled region between the
  spectrum and plus   and minus the 95\%   confidence synthetic
  spectra  in crimson, respectively.  Overall,
  the fit is good, although the residuals show that the hard SED
  over-predicts the Ly$\alpha$ line, and under-predicts the \ion{Si}{4}
  line (marked by blue arrows).  \label{fig10}}  
\end{center}
\end{figure*}

The results presented above came from simulations using a relatively
soft spectral energy distribution that may correspond to that of a
typical quasar \citep{hamann11}.  The ratios of ionic column densities
and excited states may be a function of the SED.  Therefore, we
performed an 11-bin tophat accordion model fit using {\it Cloudy}
column densities obtained when a relatively hard SED \citep{korista97}
was used.  The results are shown in Fig.~\ref{fig10}.

The reduced $\chi^2$ for the hard SED  model 
is { 1.51 (1.26), compared with a value of 1.10 (1.16) for the soft
  SED and first (second) continuum models.  As noted above, a value
  larger than 1.2 is excluded at 99\% confidence level, suggesting
  that the synthetic spectrum produced using a soft SED provide a
  significantly better fit than that produced using a hard SED.  } 
Examination of the fit residuals shows that the model slightly
over-predicts the Ly$\alpha$ line, and  under-predicts the \ion{Si}{4}
line, and those are the origins of the larger reduced $\chi^2$.  

\subsubsection{Effect of Metallicity}\label{metallicity}

Some evidence exists for enhanced metallicity in quasars
\citep[e.g.,][]{hf99,hamann02,kura02b}.  To
address potential model dependence on metallicity, we performed a
set of {\it Cloudy} runs using enhanced metallicity equivalent to
$Z=3 Z_\odot$.  Following \citet{hamann02}, all metals were set to three
times their solar value, while nitrogen was set to nine times the
solar value, and helium was set to 1.14 times the solar value.  The
relatively soft SED from \citet{hamann11} was used.  

\begin{figure*}[!t]
\epsscale{1.0}
\begin{center}
\includegraphics[width=5.5truein]{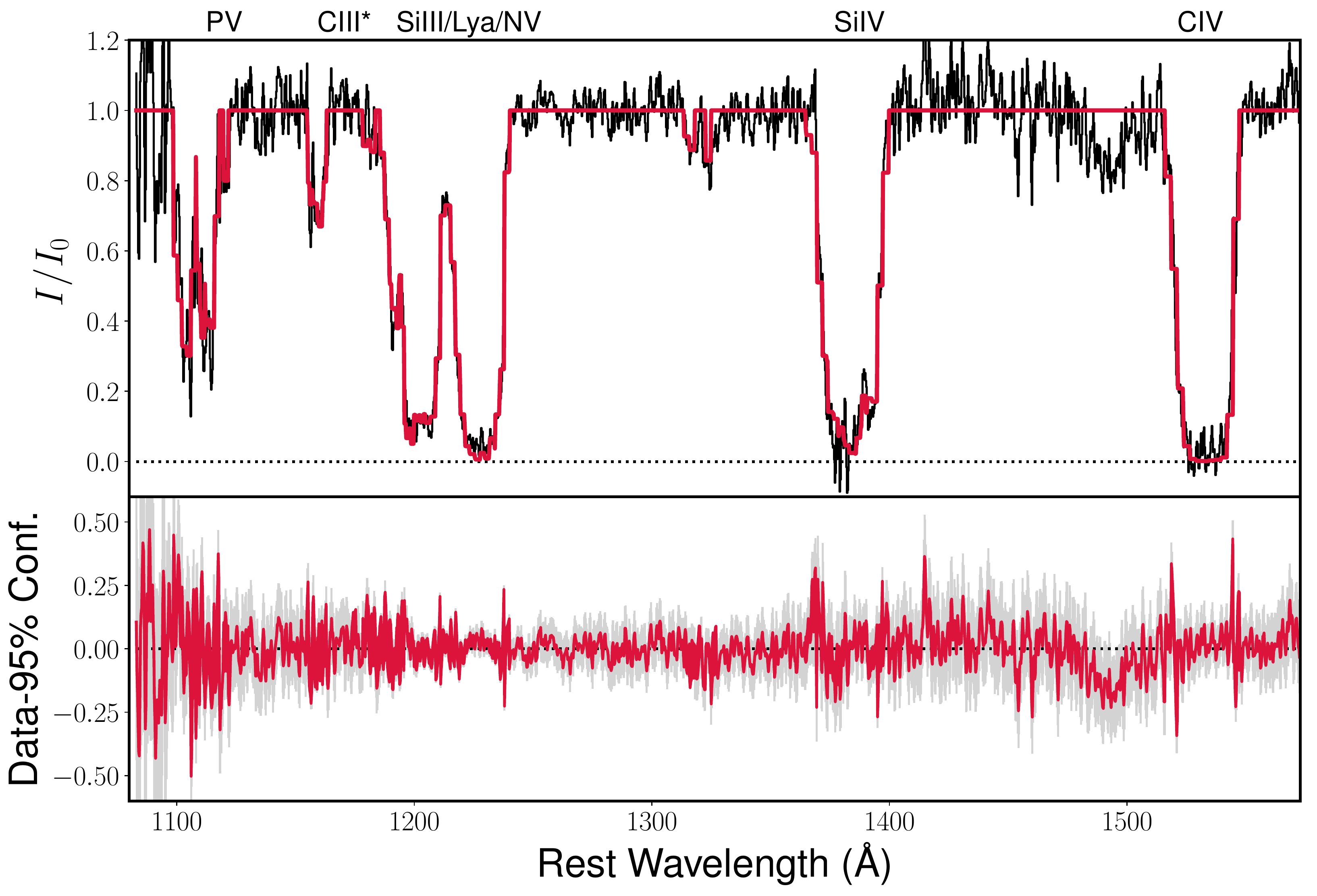}
\caption{The results from an 11-bin tophat accordion model, using
  ionic columns from {\it Cloudy} runs using enhanced metallicity
  corresponding to $Z=3 Z_\odot$. In the top panel, the  median synthetic spectrum
  (crimson) is overlaid on the continuum-normalized spectrum (black).
  The   lower panel shows the spectrum minus the median  
  model and errors in gray, and  the filled region between the
  spectrum and plus   and minus the 95\%   confidence synthetic
  spectra  in crimson, respectively.
  Overall, the higher metallicity model produces the best fit as it is
  able   to fit the \ion{Si}{4} line well without predicting too much
  Ly$\alpha$.    \label{fig11}}  
\end{center}
\end{figure*}

The results for the 11-bin tophat accordion model for the second
continuum model are shown in Fig.~\ref{fig11}.  The result is similar 
for the first continuum model.  As shown in Fig.~\ref{fig4}, the 
reduced $\chi^2$ is consistently lower than for solar
metallicity.    The fit is somewhat better in the vicinity of
Ly$\alpha$ and \ion{Si}{4}, as the model produced less neutral
hydrogen and more Si$^{+3}$ compared with the solar abundance model.
In addition, the model includes the small \ion{C}{2} lines near
1320\AA\/.   

We ran the enhanced metallicity models with 7, 8,
9, 10, 11, and 12 bins for both continua.  The  parameters as a
function of velocity are shown in Fig.~\ref{fig7} and Fig.~\ref{fig8}.
The behavior as a function of velocity is similar to the solar
metallicity case, although naturally $\log N_H -\log U$ is
systematically lower. The largest, although still subtle, difference
is  an increase in column density and stronger decrease in ionization 
parameter around $v=-2400\rm \, km\, s^{-1}$.  

As before, the column density seems to be roughly constant for
velocities more negative than $-3300\rm \, km\, s^{-1}$, with an average
column over that region of $\log N_H = 21.6$ for the metals $\times 3$
case.   The column density decreases at lower velocities, to an
approximate constant level for velocities less negative than $-2800\rm
\, km\, s^{-1}$ of  $\log N_H=20.8 \, \rm [cm^{-2}]$.

\subsection{Derived Quantities}\label{derived}

Using the results of the MCMC, we computed derived quantities such
as the total column density in the outflow, the radius of the outflow
(as inferred from the concentration represented by \ion{C}{3}*), the
mass outflow rate, the momentum flux, and the kinetic luminosity of
the outflow.  In each case, we computed these parameters individually
for each point in the MCMC chain (after cutting off the burnin), and
then extracted the median points and $1\sigma$ error bars from the
remainder.  

\begin{figure*}[!t]
\epsscale{0.7}
\begin{center}
\includegraphics[width=3.5truein]{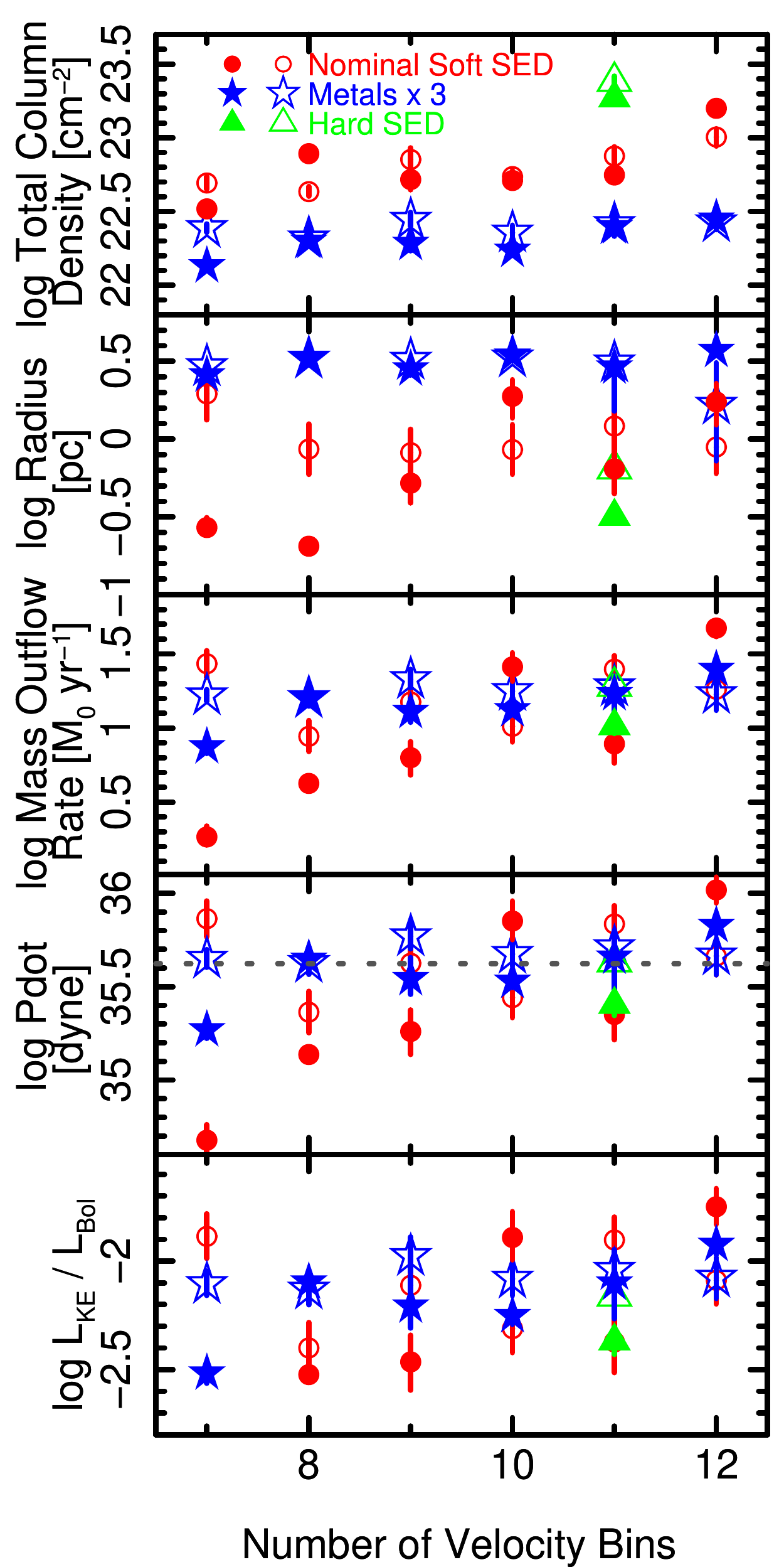}
\caption{The derived quantities from the 26  models.  In each case,
  the open (filled) symbols show the results using the first (second)
  continuum model.  The top   panel shows the
  total hydrogen column density in the outflow,   weighted by the
  covering fraction in each bin.  The    second-from-the-top panel
  shows the radius of the outflow obtained   using the   density
  constraints in the concentration.  The next   panel shows the mass
  outflow rate, assuming that the radius of all outflow  components is
  the same as the    radius of the concentration.  The fourth panel
  shows the momentum   flux, $\dot P$, applying the same
  radius   assumption, while the dashed line shows $L_{Bol}/c$.
  Finally, the   bottom panel shows the  ratio 
  of the   kinetic luminosity to the   bolometric luminosity, again
  with the   same radius assumption.  See text for further details.  In each
  case, the median and 1 sigma   error bars obtained from the MCMC model
  are   plotted.  \label{fig12}}   
\end{center}
\end{figure*}

We considered first the total hydrogen column density in the outflow.
The results are shown in Fig.~\ref{fig12}  as a function of the
number of bins, spectral energy distribution, and metallicity.  No
strong dependence on the number of bins is seen, although a larger
number of bins tends to yield a larger column density,  as the model
fits smaller-scale bumps and wiggles in the spectrum and the wings of
the lines. 

We obtained representative quantities of the derived parameters by
taking the average over the 9--12 bin cases; the 7 and 8 bin cases,
and the hard SED cases are not considered due to their
less-than-acceptable fits (\S\ref{single}, Fig.~\ref{fig4}).   

The average total column density for the nominal, rather soft SED is
$\log  N_H=22.85$ (22.91) $\rm [cm^{-2}]$ for the first (second) continuum
model.  The hard SED (for an 11-bin model) yields much larger column
density estimate, $\log N_H=23.4$ (23.3) $\rm [cm^{-2}]$.  A  harder
continuum produces a larger Str\"omgren sphere
\citep[e.g.,][Fig.\ 13]{casebeer06}, so a larger column is needed to
build up the columns of the relatively lower ionization lines such as
\ion{Si}{4} and \ion{P}{5}.   The average column density for the
enhanced metallicity case is $\log N_H=22.41$ (22.32) 
$\rm [cm^{-2}]$, about a factor of 2.2 times smaller than for the
solar metallicity.  This result makes sense, as more metal ions,
responsible for most of the absorption lines, are
available at the higher metallicity.  The
metallicity was enhanced by a factor of 3, and it is not immediately 
clear why the column density is not three times smaller than for the
solar metallicity case, although as we noted above, the best fits for
this model show a second peak in $\log N_H - \log U$ near $-2400\rm \,
km\, s^{-1}$.  It could be that the additional metals enhance the
cooling in the photoionized gas, allowing the lower-ionization lines
that constrain the column density to be present in a smaller column of
gas.  

The radius is related to the other parameters via 
$$U=\frac{\phi}{nc}=\frac{Q}{4\pi R^2 n c},$$
where $\phi$ is the photoionizing flux with units of $\rm photons\,
s^{-1}\, cm^{-2}$, and $Q$ is the number of
photoionizing photons per second emitted from the object.  The density
is constrained only within the concentration, so we compute the radius
for those velocity bins only.   We estimate $Q$ by scaling the {\it
  Cloudy} input spectral energy distribution to the observed spectrum
(corrected for redshift and Milky Way reddening), and then integrating
for energies greater than 13.6eV.  The estimate of $\log Q=56.0$ is
obtained assuming that there is no intrinsic reddening\footnote{SED
  analysis reveals no evidence for reddening in this UV-selected
  object (Paper II; Leighly et al.\ in prep.)}.   Depending on
the model, two to four bins represent the concentration.  The radius
for each simulation is taken to be the mean radius among those several 
values. The results are shown in Fig.~\ref{fig12}.   

The average inferred radius for the soft SED, solar metallicity, and
first continuum model is $\log R = -0.027 \,\rm pc$ (0.11 for the
second continuum model), corresponding to 0.94 (1.3) pc.  The average
inferred radius for the higher metallicity case is $\log R=0.51 \,\rm 
[pc]$ (0.53), or 3.2 (3.4) pc.  The difference between radius
estimates for the solar and $Z=3 Z_\odot$ models originates in a small
difference in preferred density in the concentration, being slightly
higher for the solar metallicity (average $\log n = 7.5$) than for the
higher metallicity case (average $\log n=6.2$). The reason for this
difference is not known; we speculate that it again has something to
do with the cooling of the gas.  At any rate, we can constrain the
radius of the concentration to be 1--3 parsecs.

The radius is unconstrained at high and low velocities, because these
regions, outside the concentration, are not represented in any
density-sensitive lines in the observed bandpass.  Thus the density
outside of the concentration is unbounded in the spectral
modeling.  Since the density for velocity bins higher and lower than
the concentration is unconstrained, it is consistent with the density
of the concentration.  We  therefore assumed that the whole
outflow is roughly co-spatial, and therefore assume that the inferred
radius at high and low velocities is the same as the radius for the
concentration.    

Once we made this assumption, we can compute the mass outflow
rate, given by Eq.~9 in  \citet{dunn10} \citep[see also][]{fg12}: 
$$\dot M=8\pi \mu m_p \Omega R N_H v,$$
where the mean molecular weight $\mu=1.4$, the global covering
fraction $\Omega$ is assumed to be 0.2, and we use the
covering-fraction-weighted $N_H$ discussed above.  $\dot M$ is
computed for each bin using the velocity at the midpoint of
each bin, and then summed over the whole profile.  The results are
shown in Fig.~\ref{fig12}.  The average $\log \dot M = 1.23 \, \rm [M_\odot\,
  yr^{-1}$] (1.44), or about 17
  (28) solar masses per year for the first (second) continuum
models. For the   enhanced metallicity, the 
  result was $\log \dot M = 1.28 \, \rm [M_\odot\,   yr^{-1}$] (1.22),
    about 18.9 (16.5) solar masses per year.  There is little
    difference between the solar metallicity and $Z=3 Z_\odot$ results,
    despite the fact that the column density is lower for the $Z=3 Z_\odot$
    models.  Apparently, the slightly higher values of radius for the
    $Z=3 Z_\odot$ models combined with the slightly values of the lower
    column density to produce similar outflow rate estimates.

Next we computed the momentum flux $\dot P = \dot M v$
\citep[e.g.,][]{fg12}, shown in the fourth panel of
Fig.~\ref{fig12}. If the wind  
has an optical depth to photon scattering of $\tau \sim 1$, then the
momentum flux would be on the order of the photon momentum
$L_{bol}/c$.  { In addition, the momentum flux can be used to
distinguish between momentum-conserving and energy-conserving outflows
\citep[e.g.,][]{fiore17}.}   We also plotted an estimate of $\log
L_{Bol}/c$. \citet{luo13} estimated a log bolometric luminosity of
46.1, based on scaling the \citet{richards06} template with the
observed photometry and adding a contribution based  on an estimate of
the unabsorbed X-ray emission.  We note that  essentially the same
value ($46.2$) was obtained from scaling the {\it   Cloudy} SED to the
observed continuum and integrating over it.   Using the \citet{luo13}
value, we obtain $\log L_{Bol}/c = 35.6\, \rm [dynes]$.  The
average values are $\log \dot P = 35.6 \, \rm [dynes]$ (35.8) for
solar metallicity, and  $\log \dot P = 35.7 \, \rm [dynes]$ (35.7) for
$Z=3 Z_\odot$ for the first (second) continuum models, respectively.
We found that the observed value of the momentum flux is very 
similar to the theoretical limit for a single photon scattering,
suggesting that the outflow could be photon-momentum driven. 

Finally, we computed the mechanical efficiency, the ratio of the
kinetic luminosity to 
bolometric luminosity.  The kinetic luminosity is given by
\citet[Eq.\ 11 in][]{dunn10} as $\dot E_k= \dot M v^2/2$.  The results
are shown in Fig.~\ref{fig12}.  The mean value is $\log L_{KE}/L_{Bol}
= -2.1$ ($-2.0$, excluding the 7- and 8-bin models), corresponding
to about 0.8\% (0.9\%) for the solar 
metallicity case, and $\log L_{KE}/L_{Bol}
= -2.04$ ($-2.09$), corresponding to 0.9\% (0.8\%) for the enhanced
metallicity case, for the first (second) continuum models,
respectively.  These values fall within the range of the values
required by simulations for effective feedback, 0.5\% to 5\%
\citep{dimatteo05,he10}, albeit on the low end.     

\section{Discussion}\label{discussion}

\subsection{The Black Hole Mass}\label{black_hole_mass}

As discussed above, we found that the outflow lies 1--3~pc from the
central engine.  In order to put this value into context, we estimated 
size scales in SDSS~J0850$+$4451, beginning with the black hole mass.  

To determine the radius of the broad line region, we referred to
\citet{bentz13}, who found that $\log(R_{BLR})=K+\alpha \log[\lambda 
  L_\lambda(5100)/10^{44}\rm \, erg\, s^{-1}]$. The continuum flux
density at 5100\AA\/ was estimated from the SDSS spectrum to be
$F_{5100} = 31.3 \times 10^{-17}\rm \, erg\, s^{-1}\,
cm^{-2}$\AA\/$^{-1}$.  Using the cosmological parameters used by
\citet{bentz13} 
($H_0=72\rm\,  km/s/Mpc$, $\Omega_M=0.27$,  and $\Omega_\Lambda=0.73$),
we obtained a luminosity distance $D_L=3031\,\rm Mpc$.  Using their 
best-fitting values $K=1.527^{+0.031}_{-0.031}$ and
$\alpha=0.533^{+0.035}_{-0.033}$, we obtained an estimate of the radius
of the broad-line region of  155 light days, corresponding to
$0.13^{+0.024}_{-0.021}\rm \, pc$, where the uncertainties are based
on the regression coefficient uncertainties.

SDSS~J0850+4451 has been identified as having a disk-like H$\beta$
emission-line profile \citep{luo13}.  \citet{luo13} fit the H$\beta$
line with a relativistic Keplerian disk model.  They found inner and
outer radii for the line of 450 and 4700 $r_g$ respectively.  For our
derived black hole mass, these values correspond to $r_{in}=0.035\rm                                                
\, pc$ and $r_{out}=0.37 \rm \, pc$ respectively, roughly consistent
with the \citet{bentz13} regression-estimated H$\beta$ radius of 0.13
pc. 

\begin{figure*}[!t]
\epsscale{1.0}
\begin{center}
\includegraphics[width=4.5truein]{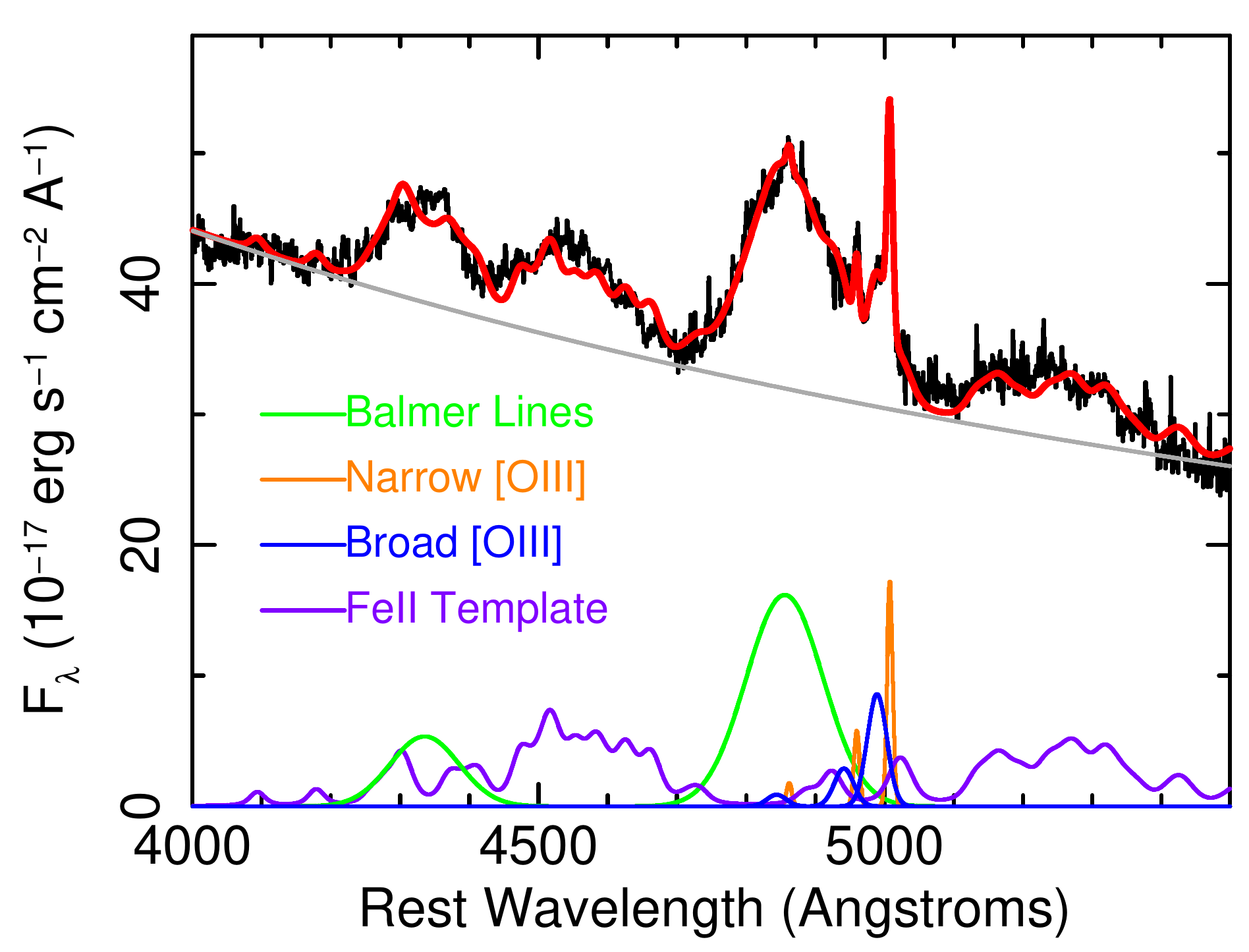}
\caption{A model of the SDSS spectrum in the region of the H$\beta$
  line.  The broad H$\beta$ line was modeled with a single Gaussian,
  yielding a FWHM of $8090 \pm 120 \rm \, km \, 
  s^{-1}$ and a black hole mass estimate of $1.6\times 10^9\rm \,
  M_\odot$. The small narrow H$\beta$ is the contribution from the 
  narrow-line region, estimated to be 10\% of the flux of
  [\ion{O}{3}]$\lambda 5007$ \citep{cohen83}.   \label{fig19}}
\end{center}
\end{figure*}

We estimated the black hole mass from the H$\beta$ line in the SDSS
spectrum in the usual way.  The data and model are shown in
Fig.~\ref{fig19}. 
We were able to obtain a good fit with a single Gaussian profile with
velocity width of $8090 \pm 120 \rm \, km\, s^{-1}$.  To estimate the
virial mass, we referred to \citet{collin06}, who provide 
line-shape-based correction factors to the FWHM-based virial product
used to estimate the black hole mass.  For a Gaussian profile,
$FWHM/\sigma_{line}=2.35$, and the scale factor for the mean spectrum
is $f=0.835$.  We estimated that the black hole mass is $1.6\times 10^9 
\rm \, M_\odot$.  With the log bolometric luminosity estimate of 46.1,
SDSS~J0850+4451 is radiating at about 6\% of the Eddington limit.

\subsection{The Location of the Outflow}\label{location}

The analysis presented in \S \ref{derived} indicated that the outflow
is located approximately 1--3 parsecs from the central engine, i.e., around the
expected size of the torus in a quasar-luminosity object.  Near-IR
reverberation has shown that the 
location of the hot inner edge of the torus is correlated with the
luminosity \citep{kishimoto07}.  We used their Eq.\ 3 to estimate that
the inner edge of the torus is $R_{\tau_K} = 0.46 \rm \, pc$, i.e.,
slightly smaller than the outflow distance.  

Another  number characterizing the torus is the  $12\, \mu \rm
m$ half-light radius.  This property does not have a clear luminosity
scaling relationship like $R_{\tau_K}$ \citep{burtscher13}.  We
estimated a plausible limit for $R_{1/2}(12\, \mu\rm m)$ by comparing
the SDSS~J0850$+$4451 bolometric luminosity with the objects in
\citet{burtscher13} Table 6.  Three objects have bolometric
luminosities within 0.2 dex of SDSS~J0850+4451, and all of these have
upper limits on their mid-IR half-light radius between 2.7 and 3.5
parsecs.  These estimates are crude, but they indicate that the
outflow is consistent with an origin near the torus in
SDSS~J0850$+$4451.  Interestingly, a torus location for the broad
absorption line outflow in the Seyfert luminosity BALQ WPVS~007 was
inferred based on variability arguments \citep{leighly15}.   

Since we have estimated the black hole mass ($1.6\times 10^9\rm \,
M_\odot$) and the radius of the outflow (1--3 parsecs), we can
estimate the escape velocity $v_{esc}=\sqrt{2GM_{BH}/R}$ at the
location of the outflow.  We find that $v_{esc}$ lies between 2100 and
$3700 \, \rm km\, s^{-1}$, interestingly close to the range of velocities
seen in the outflow.  This result suggests that the outflow could have
been accelerated from rest close to the location where it is observed,
in contrast to large-distance outflows, where the outflow velocity is
much greater than the escape velocity, and other acceleration
mechanisms such as ``cloud crushing'' \citep[e.g.,][]{fg12} are required.

\subsection{The Acceleration Mechanism}\label{mechanism}

\begin{figure*}[!t]
\epsscale{1.0}
\begin{center}
\includegraphics[width=7.0truein]{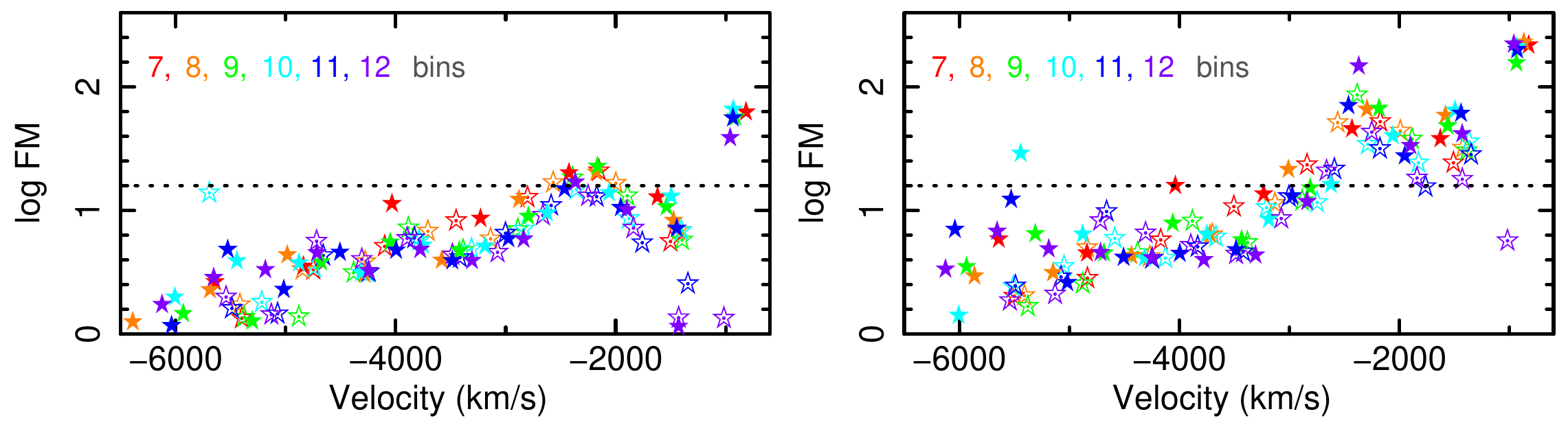}
\caption{The force multiplier computed at the MAP solution as a
  function of velocity.  The open symbols show the results for the
  first continuum model, and the solid symbols mark the solutions for
  the second continuum model.  The  left (right) panel shows the result
  for solar ($Z=3 Z_\odot$) metallicity.  The horizontal line marks
  the force multiplier value above which the flow can be radiatively
  driven.  See the text for details.   \label{fig20}} 
\end{center}
\end{figure*}

We explored the acceleration mechanism for the outflow by using
{\it Cloudy} to compute the force multiplier as a function of
velocity.  We extracted the MAP values of the ionization parameter and
column density in each velocity bin, and assumed that 
the density over the whole outflow was equal to the average MAP value
in bins representing the concentration.  The results are shown in
Fig.~\ref{fig20}.  As discussed in e.g., \citet{couto16}, the force
multiplier extracted from {\it   Cloudy} is defined as the ratio of
the total absorption cross 
section, including both line (bound-bound) and continuum (bound-free) 
processes, to the Thompson cross-section.  The absorber can be
radiatively driven if $FM \ge (L_{Bol}/L_{Edd})^{-1}$. As discussed in
\S \ref{black_hole_mass}, this quasar seems to be radiating at about
6\% of $L_{edd}$, which means that $\log FM$ should be greater
than 1.2 for the outflow to be radiatively driven.  Fig.~\ref{fig20} shows
that for the solar metallicity case, the force multiplier is generally
less than the required value, although it meets the required value for
velocities between  $-2600$ and $-1900\rm \, km\, s^{-1}$, implying
that another source of acceleration \citep[e.g., perhaps an
  MHD model,][]{kraemer18} is necessary.  On the other hand, for the
$Z= 3 Z_\odot$ case, the force multiplier exceeds the required
value significantly for velocities less than $\sim -2600\rm \, km\,
s^{-1}$, suggesting that radiative driving may be important, at least
at lower velocities in the outflow.  

The force multiplier is anti-correlated with the column
density, and the ionization parameter (see Fig.~\ref{fig7} and
Fig.~\ref{fig8}).   At larger velocities the gas may be thick but it
may be too ionized to provide sufficient opacity for radiative
driving. Alternatively, the gas may be too thick at high velocities
(especially in the region of the concentration); a very thick gas slab
is difficult to accelerate due to the loss of continuum photons by
absorption \citep[e.g.,][]{arav94a}. { This is shown by
  \citet{baskin14} in their Fig.~9.}

What is the origin of the differences between the solar metallicity
case and the $Z=3 Z_\odot$?  We expect a larger force
multiplier for a higher metallicity, as metals provide the bulk of the
scattering opacity, as observed. The simulations also show that that a
larger fraction of the acceleration in the higher metallicity case is 
attributed to bound-bound interactions.  However, the  offset between
the $\log FM$ values is not constant between the two metallicity
cases, suggesting a contribution from differences in the ionization
state of the gas and column density as well.   

{ The outflow in SDSS~J0850$+$4451 originates near the torus.  This
suggests that dust could play a role in the wind acceleration. The
equivalent dust cross section to scattering is 500--1000 times the
electron scattering cross section for a typical quasar SED, which
means that a typical quasar is super-Eddington with respect to dust,
and may imply that dust-driven outflows can contribute to feedback 
\citep[e.g.,][and references therein]{fabian12, roth12}.  Some models
for the torus that take into account radiation-driven outflows find
that a strong wind is produced \citep[e.g.,][]{kk94,gallagher15}.  For
example, \citet{ck16} predict a wind along the inner edge of the torus
with velocity $\sim 5000 (M/10^7M_\odot)^{1/4}
[L_{UV}/(0.1L_{Edd})]^{1/4} \rm \, km\, s^{-1}$ that carries $\sim 0.1
(M/10^7 M_\odot)^{3/4} [L_{UV}/(0.1   L_{Edd}]^{3/4} \rm M_\odot\,
yr^{-1}$ where $M$, $L_{UV}$, and $L_{Edd}$ are the black hole mass,
UV luminosity, and Eddington luminosity respectively.  Recent models
of the broad line region also appeal to radiation pressure on dust
\citep{ch11,czerny17,bl18}. Thus, it seems that radiative acceleration
on dust may provide more than enough energy to power winds that
originate in the vicinity of the torus and perhaps beyond.

The difficulty with this scenario is that while BALQ spectra are
typically more reddened than quasars without broad absorption lines,
only a small fraction show large amounts of reddening \citep[e.g., 13\% show
$E(B-V) > 0.1$ and 1.3\% show $E(B-V) > 0.2$,][]{krawczyk15}.
SDSS~J0850$+$4451, being UV selected, shows no evidence for
reddening.  In contrast, for a standard dust-to-gas ratio
\citep{bohlin78}, a log hydrogen equivalent column density of 22.9
predicts $E(B-V)=13.7$, far too large to be realistic.  So the dust
must be separated from the gas.  Dust is bound to the gas by
collisions \citep{wick66}, and that mechanism becomes inefficient at
low densities, allowing the dust to drift.  Evidence for this
mechanism is found in AGB stars \citep[][and references therein]{ho18}.
We speculate that it is conceivable that the dusty wind is
accelerated from the vicinity of the torus, and when it reaches a
certain density, the dust continues to be accelerated, perhaps
ultimately forming a scattering halo that may be observed as a polar
outflow \citep{honig13, hk17}, or be responsible for polarization in
BALQs \citep{ogle99} and red quasars \citep{alexandroff18}.  The gas,
which is left behind, would still have the momentum imparted during
the dust acceleration phase, and could then be responsible for the
broad absorption lines.   
}

\subsection{Where Does SDSS J0850$+$4451 Fit In?}\label{context}

In this paper, we have performed a detailed analysis of the absorption
lines in SDSS~J0850$+$4451.  In this section, we compare
SDSS~J0850$+$4451 with other BALQs in order to gauge how typical
this quasar is.

SDSS~J0850$+$4451 was selected for observation using {\it HST} COS from a
small sample of SDSS LoBAL quasars for which we had observations of
\ion{He}{1}*$\lambda 10830$ using either Gemini GNIRS and/or LBT
LUCI. Our intention was to compare the optical depths of
\ion{He}{1}*$\lambda 10830$ and \ion{P}{5} to investigate the nature
of partial covering; this is discussed in Paper II (Leighly et al.\ in
prep.).   We used {\em GALEX}
to ensure that the target objects would be bright enough for {\it HST}
to obtain a good signal-to-noise ratio in a reasonable exposure time.
Thus, SDSS~J0850$+$4451 is relatively blue.   { SED fitting, to be
presented in Paper II (Leighly et al.\  in prep.) reveals no evidence
for significant intrinsic reddening.}  In contrast, BALQs tend to
be reddened compared with the normal quasar population, although many
BALQs with little reddening are found \citep[e.g.,][]{krawczyk15}.
Indeed, SED fitting of the optical and IR photometry shows that the
torus emission is relatively weak (Paper II, Leighly et al.\ in
prep.) suggesting that SDSS~J0850$+$4451 might be relatively dust-free 
altogether. 

Despite the lack of reddening, SDSS~J0850$+$4451 was found to be X-ray
weak in a {\it Chandra} observation \citep{luo13}. Only three hard
photons, all with energies greater than 6.2 keV in the rest frame were  
detected.  Assuming a typical quasar X-ray spectrum, \citet{luo13}
found that a column of  $N_H \approx 7 \times 10^{23}\rm \,
cm^{-2}$ was necessary to produce three hard photons and no soft
photons. 
BALQs are known to be X-ray weak, and LoBAL quasars are known to be 
generally significantly X-ray weaker than high-ionization BALQs
\citep[e.g.,][]{green01, gallagher02b,gallagher06}.  Generally, this
X-ray weakness is inferred to be due to absorption.  Sometimes the
column densities can be measured directly from the X-ray spectrum
\citep[e.g.,][]{gallagher02}, but often, only a few photons are
detected, and the attenuation in the X-ray band compared with the
optical band is assumed to originate from absorption, and in that
case, the column density can be estimated. \citet{luo14} (their
Fig.~4)  shows that the $N_H$ estimated for SDSS~J0850$+$4451 is
relatively large compared with other BALQs.  Alternatively, some BALQs
have been shown to be intrinsically X-ray weak
\citep[e.g.,][]{luo14}. That may not be the case for SDSS~J0850$+$4451
as it shows relatively typical UV emission lines and ratios, in
comparison with the weak line emission in the intrinsically X-ray weak
quasar PHL~1811 \citep{leighly07a, leighly07}.  Nevertheless, it
appears that SDSS~J0850$+$4451 is  typical LoBAL in its X-ray properties.  

The broad absorption lines in SDSS~J0850$+$4451 have a maximum
velocity of  $\sim -5500\rm \, km\, s^{-1}$ and a minimum
velocity of  $\sim -1400\rm \, km\, s^{-1}$, and therefore a width of
about  $4000\rm \, km\, s^{-1}$, and a middle velocity offset of $\sim
-3500\rm \, km \, s^{-1}$.   This velocity width appears to be rather
typical of BALQs \citep{baskin15}.  Several investigators have observed
a rough upper envelope of maximum velocity with optical luminosity
\citep{laor02, ganguly07}.  For $H_0=70\rm \, km\, s^{-1}\, Mpc^{-1}$,
$\Omega_M=0.3$, $\Omega_\Lambda=0.7$, $\log \lambda L_\lambda$ at
3000\AA\/ is 45.4 [$\rm erg\, s^{-1}$]. Fig.\ 7 in \citet{ganguly07}
shows that a  
maximum outflow velocity of $5500 \rm \, km\, s^{-1}$ appears to be
consistent with the average for an object with this luminosity.  

As discussed in \S \ref{black_hole_mass}, the rather broad Balmer line
observed in SDSS~J0850$+$4451 yields a large black hole estimate of
$1.6 \times 10^9 \rm \, M_\odot$, and for a bolometric luminosity
estimate of $46.1 \, \rm  [erg\, s^{-1}]$ \citep{luo13}, the object is
radiating at only 6\% of 
Eddington. This value is low for a type 1 object.  In contrast,
\citet{yw03} observed $z\sim 2$ BALQs in the infrared band and found
high (of order $\sim 1$) Eddington ratios. They noted that this might
be a selection effect due to observing the brightest objects; on the
other hand, they found that most objects had strong ``Eigenvector 1''
line emission patterns including very strong \ion{Fe}{2} and weak
[\ion{O}{3}], also an indication of a high Eddington ratio.  In
contrast, SDSS~J0850$+$4451, with its broad Balmer lines and modest
\ion{Fe}{2} emission has 
emission line properties mostly consistent with a low (for a Seyfert
1) accretion rate.  The [\ion{O}{3}] appears too weak for an object
with such a broad H$\beta$ line, but weak [\ion{O}{3}] seems to be
common among LoBAL quasars \citep[e.g.,][and references
  therein]{schulze17}.  Empirically, it has been 
suggested that BALQs are typically high Eddington objects
\citep[e.g.,][]{boroson02}, and this has also been suggested on
theoretical grounds \citep{zk13}.  SDSS~J0850$+$4451's low Eddington
ratio would seem to make it anomalous, because, as shown in
\citet{ganguly07} (their Fig.~6), very few of the \citet{trump06}
BALQs radiate at  less than 10\% Eddington.   However, a more recent
study by \citet{schulze17} revealed no difference in black hole mass and
Eddington ratio between a sample of 22 LoBAL quasars and unabsorbed
objects, and several of their $z \sim 0.6$ sample showed Eddington
ratios less than 10\%.  Thus general claims that all LoBALQs are high
Eddington-ratio objects do not seem justified, although high
Eddington-ratio objects may be over-represented in this population.

SDSS~J0850$+$4451 has a kinetic-to-bolometric luminosity ratio for the
broad absorption lines of 0.8--0.9\% (Fig.~\ref{fig12}).  We note
that there is also a blueshifted component of the [\ion{O}{3}] line,
modeled as an additional Gaussian with velocity offset of $-1150\rm \,
km\, s^{-1}$ (\S~\ref{black_hole_mass}), although we have no
information about the spatial extent of this emission.
\citet{fiore17} attempt to bring together a compendium of outflow
indicators.  While their information 
is incomplete for BALQ measurements, their results are
nevertheless useful for comparison.  Our $\dot E_{kin}/L_{bol}$ ratio
is comparable to other BALQs and ionized winds for objects of the same 
bolometric luminosity.  Like the other BALQs, our $v_{max}$ lies
between the relatively low-velocity ionized gas and molecular outflows,
and the ultra-fast outflows (UFOs).  Our momentum flux ratio $\dot
P_{OF}/\dot P_{AGN}$ is approximately 1, and is typical of ionized
winds and UFOs, and less than the molecular outflows.  

The outflow in SDSS~J0850$+$4451 is located 1-3 parsecs from the
central engine, consistent with the estimated location of the
torus. Interestingly, a similar location was inferred for the broad
absorption line outflow in the Seyfert-luminosity Narrow-line Seyfert
1 Galaxy WPVS~007 based on variability arguments \citep{leighly15}.   
Density-constrained distances have been measured for a handful of
objects, and these span a wide range, from the vicinity of the torus
(parsec scale) to kiloparsec scale \citep[e.g.,][]{lucy14, dabbieri18,
  arav18}.  In comparison with the kiloparsec-scale outflows, the
outflow in SDSS~J0850$+$4451 appears to be relatively compact.   

The discussion and comparisons above indicate that SDSS~J0850$+$4451
has an outflow characterized by typical offset velocity and velocity
width.  But compared with other BALQs, it radiates at a relatively low 
Eddington ratio, has relatively broad emission lines, and has
relatively low reddening and weak torus emission that suggest a low
dust content.  The outflow is observed near the torus, rather than at
kiloparsec distances as has been found in some other BALQs, and
therefore seems relatively compact.  Although a conclusive comparison
will have to wait until we have analyzed more objects, we suggest that
the feedback interaction between the quasar nucleus and the host
galaxy is not currently ongoing, although it may have been in the
past.  Indeed, SED fitting gave only an upper limit on the
star-formation rate \citep{lazarova12}. Nevertheless, the 
outflow SDSS~J0850$+$4451 hosts is likely to be important to the
operation of the central engine.  The accretion rate is estimated to
be only $2.2 \rm \, M_\odot\, yr^{-1}$, while the outflow rate is
about 8 times higher.  So if this object did not host an outflow, it
might accrete at a higher rate, and the central engine and emission
line  properties might be  much different. 

\section{Summary and Future Prospects}\label{conclusions}

\subsection{Summary of Results}

In this paper, we present a detailed analysis of low redshift LoBAL
quasar SDSS~J0850$+$4451, using the novel spectral synthesis code
{\it SimBAL}.  Our principal results follow.

\begin{enumerate}
\item We introduced the {\it SimBAL} analysis method (\S~\ref{simbal}).
  Using large grids of ionic column densities extracted from {\it
    Cloudy} models, we created synthetic spectra as a function of
  velocity, covering fraction, ionization parameter, density, and a
  combination column density parameter $\log N_H - \log U$.  A
  forward-modeling 
  spectral-synthesis approach was then used to compare the
  continuum-normalized {\it HST} spectrum with the synthetic spectra
  enabled by the Markov Chain Monte Carlo code {\tt emcee}.  The
  results included best-fitting spectra and  posterior probability
  distributions of model   parameters from which values and limits on
  physical parameters were  extracted. 
\item We investigated the systematics of our method by using two
continuum models, two spectral energy distributions, a range of the
number of velocity bins, and two values of the metallicity. { Most
  of these combinations fit the data relatively well (Fig.~\ref{fig6},
  \ref{fig6b}, \ref{fig10}), although the   models with the smallest
  number of bins and the models using the   hard spectral energy
  distribution are not favored statistically   (Fig.~\ref{fig4}).    }
\item Our models revealed interesting structure
  as a function of   velocity (Fig.~\ref{fig7},\ref{fig8}).  The
  density-sensitive line   \ion{C}{3}* appears in the   spectrum over
  a limited velocity   range, and in this region, $\log   N_H-\log U$
  is larger than at   other velocities.  There is evidence   that both
  $\log U$ and $\log   N_H$ are larger at higher velocities. 
  There is a strong decrease in covering fraction with velocity.
\item We were able to extract robust estimates of the total column
  density of the gas depending on the metallicity: 22.9 [$\rm cm^{-2}$]
  for solar, and 22.4 [$\rm cm^{-2}$] for $Z=3 Z_\odot$. The 
  density-sensitive line \ion{C}{3}* line indicated a distance from
  the central engine of 1--3 parsecs. Assuming that all the gas lies 
  at approximately   the same distance from the central engine, we
  found that the mass outflow rate is  17--28 solar masses per year,
  the log of the   momentum flux is  35.6--35.8 [dynes], consistent
  with $L_{Bol}/c$,   and the ratio of the kinematic to bolometric
  luminosity is around   0.8--0.9\%.   
\item Using these results, we built a physical picture of the outflow
  in SDSS~J0850$+$4451.  The outflow location based on the gas density
  is consistent with an origin in the torus, where the escape velocity
  is interestingly close to the observed velocities in the outflow.
  Force multiplier analysis indicates that at least the lower velocity
  portions of the outflow might be consistent with acceleration by
  radiative line driving along our line of sight, and we speculated
  that dust scattering may also play a role, although selection for
  {\it     HST} observation means 
  that SDSS~J0850$+$4451 is a relatively blue object lacking the
  reddening that is common in BALQs in general and LoBALQs
  specifically.   SDSS~J0850$+$4451 has an Eddington ratio of just 6\%,
  lower than that of the general population of BALQs.  Given the
  compact nature of the outflow, we speculated  that
  SDSS~J0850$+$4451 is past the   era of feedback, if it occurred
  previously.  A JWST study of its host galaxy to determine whether it
  is quiescent or star-forming would be an interesting follow-up.
  Nevertheless, we contend that   the outflow is of integral
  importance to the nature of the central   engine, since the outflow
  rate is estimated to be nearly an order of   magnitude greater than
  the accretion rate. 
\end{enumerate}

\subsection{The Future}\label{future}

{\it SimBAL} has enabled us to perform an analysis of the rest-frame
UV broad absorption line outflow in the low-redshift
quasar SDSS~J0850$+$4451 that is unprecedented in detail.  Clearly,
however, the real 
power of {\it SimBAL} will be manifest when we compare SDSS~J0850$+$4451
with other objects.  For example, it will be interesting to compare with
analysis of SDSS~J142927.28$+$523849.5, a second object observed as
part of our {\it HST} program, that shows strong Eigenvector-1 
properties, i.e, a narrow H$\beta$ line and strong \ion{Fe}{2}
emission, likely indicating a high accretion rate (Leighly et
al.\ 2018, in prep.).  But that is only the beginning.  Large samples
from SDSS/BOSS can be analyzed to infer the general properties of
outflows.  It should be emphasized that these are {\it physical}
properties of the gas, rather than the empirical  properties (e.g.,
balnicity, maximum depth, width, maximum velocity) that dominate BALQ
studies today.  We will be able to, for example, determine ionization
parameters, column densities, and covering fractions in a large number
of objects.  Densities and outflow radii will be extracted from a
subset.  For example, analysis of a sample of FeLoBALs has revealed
outflow radii that span four orders of magnitude, and a general lack
of broad lines at kpc scales \citep[][Leighly et al.\ in
  prep.]{dabbieri18}.  We will be able to look for trends as a
function of velocity among outflows; for example, perhaps the covering
fractions usually decrease with increasing velocity.  We will be 
able to learn whether outflow concentrations such as the one
observed in SDSS~J0850$+$4451 are common, or whether the gas
properties are more uniform generally.  We believe that {\it SimBAL} will 
be able to revolutionize the study of broad absorption line quasars.  

To reach the full potential of the approach, several improvements are
being implemented.  An important one is that we need to account for 
systematic uncertainties due to continuum model and placement.  We are
developing a principal components analysis approach that will allow us
to simultaneously model the absorption and the continuum
\citep{marrs17,wagner17}, which has already been implemented in the
near-UV \citep[][Leighly et al., in prep.]{leighly_aas17,dabbieri18}.
The far UV is a harder nut to crack, given the Ly$\alpha$ forest
contamination, and we are currently testing some promising approaches
(Choi et al.\ 2018, in prep.). Other improvements, either implemented
or planned, include schemes to speed up the code, reduce the memory
requirements, and implement various automated checks on the solutions,
as well as an interactive interface that can be used to generate
starting points.  We plan to release the software to the community
once it is fully vetted.   

\acknowledgements

KML thanks Chris Done for suggesting looking at the force multiplier,
{ Coleman Krawczyk for advice on using {\tt emcee}, and Bozena
  Czerny for a suggestion regarding dust and gas decoupling.}
KML acknowledges useful discussions with Mike Eracleous and with the
current {\it SimBAL} group: Joseph Hyunseop Choi, Collin Dabbieri, Amy 
Griffin, Francis MacInnis, Adam Marrs, and  Cassidy Wagner.  {
  {\it SimBAL} was conceived during   KML's sabbatical leave, and she
  acknowledges the University of   Oklahoma and the Homer L.\ Dodge
  Department of Physics and Astronomy   for support of her research.}
{ Some of 
the computing for this project was performed at the OU Supercomputing
Center for Education \& Research (OSCER) at the University of Oklahoma
(OU).}    Support   for program 13016  was provided by NASA through a
grant from the Space Telescope Science Institute, which is operated by
the Association of Universities for Research in Astronomy, Inc., under
NASA contract NAS 5-26555.  Support for {\it SimBAL} development was 
provided by NSF Astronomy and Astrophysics Grant No.\ 1518382.  DT
acknowledges the Homer L.\ Dodge Department of Physics and Astronomy
of the University of Oklahoma for graciously hosting his sabbatical
visit in 2017.   SCG acknowledges the Natural Sciences and Engineering
Research Council of Canada.

\facility{HST (COS)}

\software{emcee \citep{emcee}, Cloudy \citep{ferland13}}

\appendix

\section{Potential Systematic Effects}\label{systematic}

In this paper, we build upon current state-of-the-art methods for
quantitative comparison of spectra with photoionization models.  In 
these methods, the photoionization models are constructed assuming a
constant density and dust-free slab of gas \citep[e.g.,][]{moe09,
  dunn10, arav13, borguet12, borguet13, lucy14,chamberlain15,xu18}.  
It is possible that those assumptions can be tested with SimBAL.  In
this section, we perform some limited tests; a more comprehensive
discussion is beyond the scope of this paper.  We use as an example
one of our solutions, an 11-bin model using the nominal soft spectral
energy distribution and solar abundances fit to the spectrum
normalized by the first continuum model.

\subsection{Dust}

The inferred location of the absorbing gas in SDSS~J0850$+$4451 is
1--$3\rm \, pc$ (\S~\ref{derived}).  This distance is larger than the
dust sublimation radius, estimated to be about 0.2~pc based on the
inferred bolometric luminosity \citep{ld93}.  It therefore may be 
expected that dust would be present in the absorbing outflow.  Dust
dramatically alters the photoionized ionic column densities due to
attenuation of the continuum and metal depletion.  As
discussed in \S\ref{context}, and discussed in more detail in
Paper II (Leighly et al.\ in prep.), SDSS~J0850$+$4451 was selected
from among our \ion{He}{1}* BALSQSOs
because {\em GALEX} photometry showed that it bright enough in the UV
to observe using {\it HST} in a reasonable amount of time.   Thus,
SDSS~J0850$+$4451 is a rather blue object, and spectral energy
distribution modeling of the broad band photometry indicates that
reddening is negligible (Paper II; Leighly et al.\ in prep.).  There
is therefore no evidence for dust in the absorber in
SDSS~J0850$+$4451.   

The recent literature indicates that dust seems to be quite
complicated in quasars and AGN.  In several objects, anomalously steep
and unusually-shaped reddening curves have been found
\citep[e.g.,][]{leighly09, jiang13, leighly14}.  In Mrk~231, we found that the
unusual reddening curve was similar to those previously used to
explain the low values of total-to-selective extinction in Type Ia
supernovae \citep{leighly14}.  While SMC extinction generally provides
a good fit to quasar photometry \citep{krawczyk15}, \citet{zafar15}
find extinction curves for heavily reddened quasars that are steeper
than the SMC, although their results may depend somewhat on the
assumed shape of the intrinsic continuum \citep{collinson17}.  They
speculate that large dust grains might be destroyed by the active
nucleus.  \citet{gallerani10} found that the extinction laws for high
redshift quasars deviate significantly from the SMC, being flatter in
the UV.  This trend was particularly pronounced for BALQs.  They
suggest that the difference between high  and low redshift
quasars may originate in the fact that the relative contribution of
AGBs and SNes in quasars is strongly dependent on the star formation
history and the age of the Universe. 

Nonetheless, the question of the presence of dust in outflows is
interesting. \citet{leighly14} explored dust and depletion in some
detail in Mrk~231, a heavily reddened object in which the presence of
dust is undeniable.  We found that dust was necessary to produce the
\ion{Na}{1} absorption line without requiring an unreasonably large
column density.   We also found evidence for density enhancement in
the partially ionized zone (from, e.g., a shock).  While it was
possible to perform a thorough exploration of parameter space for a
single object, it is not feasible to include so many independent
parameters in {\it SimBAL}.  In the future, as appropriate, we may
attempt to analyze the effect of dust, perhaps in a binary way, as we
have explored the dependence on metallicity and spectral energy
distribution in this paper.

\subsection{Filtering}

Our tophat opacity model splits the absorption profile into adjacent
velocity bins, allowing us to derive physical parameters of the
outflow as a function of velocity.  This information may ultimately
help us constrain the origin and acceleration of the outflows.  Biased
by the fact that we observe only the radial component of the outflow,
it is tempting to interpret the tophat model physically, i.e., to
assume that we are seeing an accelerating outflow, with the lowest
velocity slab closest to the continuum source, and higher velocity
slabs sequentially behind one another.  If this were the case, then
one would expect outer slabs to be illuminated by the transmitted
continuum of inner slabs. Alternatively, the flow could be
decelerating. 
The transmitted continuum would be deficient in photons of certain
energies, depending on the ionization parameter and thickness of the
bin.  Filtering may play a role in producing the characteristic
intermediate-ionization emission lines in Narrow-line Seyfert 1
galaxies \citep{leighly04}, and has been used to explain the lack of
the \ion{He}{1}$\lambda 5876$ emission line in weak-line quasar
PHL~1811 \citep{leighly07}. It is not clear whether filtering plays
an important role in BAL outflows.

\begin{figure*}[!t]
\epsscale{1.0}
\begin{center}
\includegraphics[width=7.0truein]{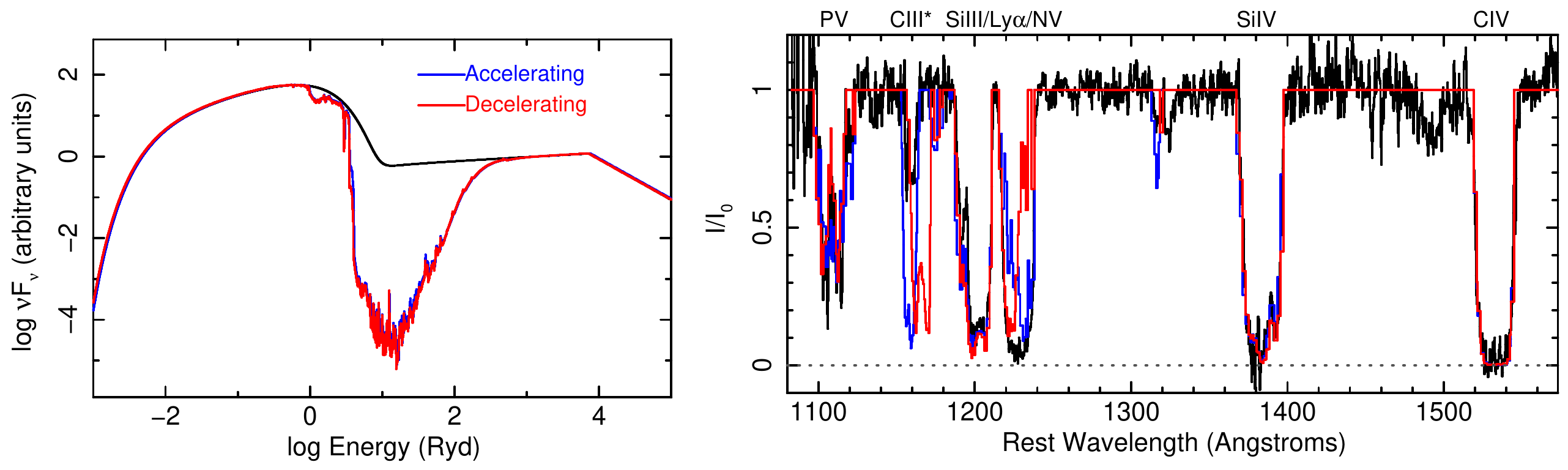}
\caption{Tests of the effect of the effect of filtering.
  {\it Left:} The filtered continua for the accelerating (blue) and
  decelerating (red) cases, compared with the unabsorbed continuum
  (black), computed as described in the text. The filtered continua 
  are not much different from the unabsorbed in the hydrogen continuum
  (0--4 Rydbergs), but are much weaker in the helium continuum
  ($> 4\rm \, Rydbergs$).  {\it Right:}  
The SDSS~J0850$+$4451 spectrum overlaid with the
  best-fit 11-bin nominal SED solar abundance model subject to
  filtering, as described in the text.  The signature of acceleration
  (blue) and deceleration (red) are strongly seen in \ion{N}{5}
  absorption line, with the decrease in opacity at high and low
  velocities,   respectively.  \label{filtered}}     
\end{center}
\end{figure*}

Starting with the 11-bin model illuminated by the soft spectral energy 
distribution and solar abundances, we tested both acceleration and
deceleration scenarios, as follows.  For acceleration, we first
redshifted the illuminating continuum by the offset velocity of the
lowest-velocity bin.  We ran the {\it Cloudy} simulation, and
extracted the total (i.e., including diffuse emission) transmitted
continuum. To take the covering fraction into account, we computed the
inferred opacity as a function of energy (the continua are represented in
Rydbergs) and used the power-law partial covering model to compute the
inferred $I/I_0$.  Multiplying by the input continuum gave the
covering-fraction weighted transmitted continuum.  We redshifted this
continuum to account for the velocity offset of the next bin, and
illuminated the second bin, harvesting the transmitted continuum, and
so on through the eleven bins.  The 
resulting continua transmitted through the whole outflow for the
accelerating and decelerating cases are shown in
Fig.~\ref{filtered}.  Since the simulation is matter
bounded\footnote{A matter-bounded slab is optically thin to the
  hydrogen continuum, while an ionization-bounded slab is optically
  thick to the hydrogen continuum.}, we do
not lose much light in the hydrogen continuum; the ionization
parameter $U$ (computed from 1 Rydberg to higher energies) decreases
by only 50\%.  In contrast, the decrease in the helium continuum is
dramatic; ionization parameter $U$ computed for $E>4\rm \, Rydbergs$
is lower by a factor of 150.

The resulting synthetic spectra are shown in Fig.~\ref{filtered}.
The \ion{C}{4} and \ion{Si}{4} are largely unaffected.  This is
expected since the hydrogen continuum is not much altered by filtering
and the ionization potentials to create C$^{+3}$ (47.9~eV) and
Si$^{+3}$ (33.5~eV) are both less than 4 Rydbergs (54.4 eV).  In
contrast, \ion{N}{5}, with ionization potential to create
N$^{+4}=77.5\rm \, eV$, is strongly affected.  The signature of
acceleration and deceleration are clearly seen, from the decrease of
\ion{N}{5} opacity at high and low velocities, respectively.  

The other large change in the synthetic spectrum is in the \ion{C}{3}* 
line.  As seen in Fig.~\ref{fig9b}, this line is sensitive to column
density. For a fixed $\log N_H-\log U$, with the ionization parameter
effectively decreasing as the continuum becomes more and more
filtered, the slab produces more lower-ionization lines at the back
end than before. 

This result, although limited in generality, is very instructive
nonetheless.  If filtering is typically important in BAL outflows, we  
would expect to measure a consistent increase or decrease in
ionization with velocity.  Generally speaking, we do not see that.
High-ionization lines are usually the broadest of all, encompassing
low-ionization lines in velocity space.  For example, \ion{O}{6} lines
are often inferred to be very broad \citep[e.g.,][]{leighly09}.
Instead, in the case of SDSS~J0850$+$4451, we find that the ionization
parameter varies only subtly with velocity (Fig.~\ref{fig6b} and
Fig.~\ref{fig7}).  

It therefore seems that gas at all velocities is illuminated by the
continuum from the central engine.  This may mean that the direction
of the outflow is oblique to the line of sight
\citep[e.g.,][Fig.~3]{arav04}, so gas at all velocities have a clear
view of the nucleus.  It is also not clear what the role of partial
covering is in the question of filtering.  By definition, gas that
partially covers the continuum leaves part of the line of sight free
from obscuration so that the continuum can illuminate gas at larger
radii free from attenuation.

\subsection{Constant Pressure}

\citet{baskin14a} suggested that radiation pressure compression
(confinement) may apply to broad absorption line outflows, although
they did not perform a quantitative comparison with spectra.  Radiation
pressure confinement was first applied to the narrow-line region in
AGN by \citet{dopita2002}.  This model has been subsequently applied
to the extended narrow line region by \citet{stern14}, to the broad
line region by \citet{baskin14}, and to warm absorbers by
\citet{stern14a}.  This model assumes constant total pressure in the
photoionized slab, rather than constant density. 

We tested this scenario by running constant pressure models using
the best-fit parameters for the 11-bin soft SED solar abundances
model.  We did not expect dramatic differences between the constant
density and constant pressure models for SDSS~J0850$+$4451 because the
outflows are optically thin to the Hydrogen continuum, and are therefore
matter bounded rather than ionization bounded.  Our previous
experience with constant pressure and compression in the context of
Mrk~231 \citep{leighly14} suggested that many of the differences
between constant density and constant pressure occur near the hydrogen
ionization front and in the partially ionized zone.  As discussed
\citet{stern16}, in the context of emission line models, gas can be
most effectively compressed if ionization bounded (rather than
matter bounded, as in SDSS~J0850$+$4451) so that there is a slab of
mostly neutral material against which the ionized gas can be
compressed.    

\begin{figure*}[!t]
\epsscale{1.0}
\begin{center}
\includegraphics[width=7.0truein]{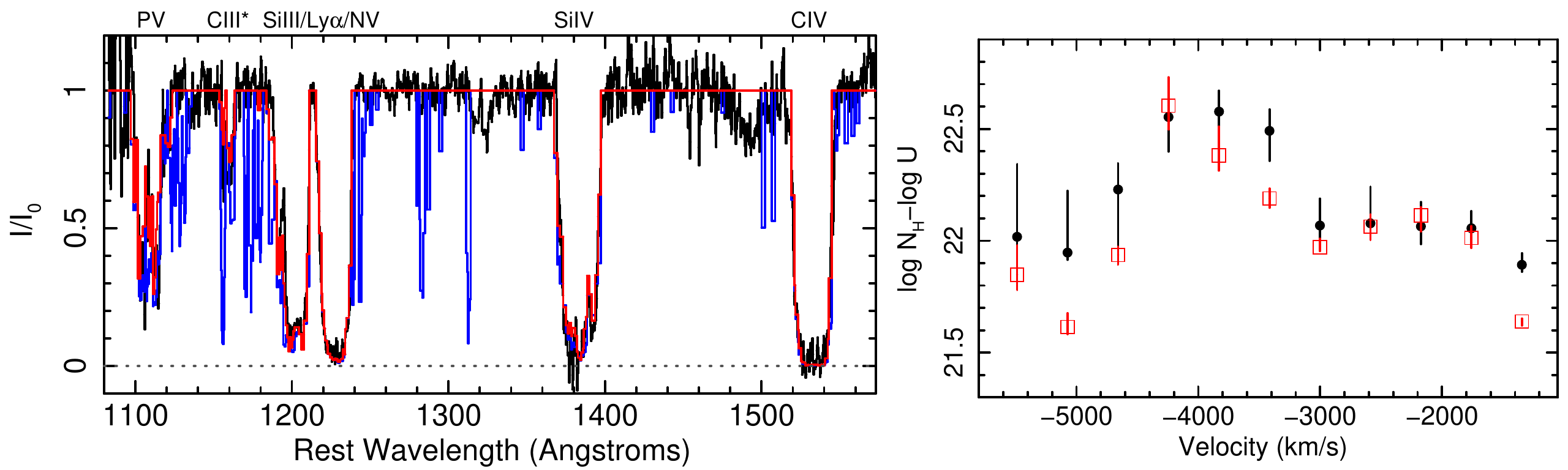}
\caption{Tests of the effect of the constant pressure assumption.
  {\it Left:} The SDSS~J0850$+$4451 spectrum overlaid with the
  best-fit 11-bin soft SED solar abundance model with the total
  pressure constrained to be constant (blue line).  Comparison with
  Fig.~\ref{fig9b} shows that the $\log N_H - \log U$ appears to be
  too high.  The red line shows the same model with $\log N_H -\log U$
  allowed to vary, yielding an acceptable fit. {\it Right:}  The best
  fitting $\log N_H-\log U$ parameters for the best fit shown in
  Fig.~\ref{fig6} (black), and the best-fitting constant pressure
  values (red).  The constant pressure model indicates a slightly
  lower column density.    \label{constant_pressure}}    
\end{center}
\end{figure*}

The blue line in Fig.~\ref{constant_pressure} shows this model
overlaid on the data. The appearance of strong low-ionization lines
such as \ion{C}{2}$\lambda 1335$ that are not observed, and comparison
with Fig.~\ref{fig9b} (especially the accompanying animations)
suggests that the column density is slightly too high.  We tested this
idea by performing an MCMC simulation, allowing the eleven values of
$\log N_H - \log U$ to be free, and fixing the other parameters at
their best-fitting values.  The red line shows that the resulting
simulated spectrum agrees very well with the observed spectrum.  The
reduced $\chi^2$ is 1.28, higher than the value for the best fitting
constant density model of 1.11.  Undoubtedly, a better fit could be
obtained if all parameters were allowed to vary.  That experiment is
beyond the scope of this paper, and may be interesting to pursue in
future work.

The log of the covering fraction-weighted  column density from
the constant density best fit was $22.88\pm 0.06$ [$\rm cm^{-2}$] (95\%
uncertainties).  For the constant pressure and variable $\log N_H-\log
U$ model, the log of the covering fraction-weighted column density was
$22.71^{+0.04}_{-0.03}$  [$\rm cm^{-2}$].  It is expected that the
total column density should be slightly lower, as observed, since in a
constant pressure model, the density increases into the slab and
therefore the more highly ionized gas is found closer to the
illuminated face.  The difference in total hydrogen column density in
this case is very small, as expected.

\end{document}